\documentclass[showpacs, amsmath,amssymb, aps]{revtex4}

\usepackage{tabularx}
\usepackage{graphicx}
\usepackage{float}
\usepackage{dcolumn}
\usepackage{bm}
\usepackage{mathrsfs}
\usepackage{amsmath}

\begin{document}


\title{Relevant scales for the $C$-metric with positive cosmological constant}
\author{D. Batic}
\email{davide.batic@ku.ac.ae}
\affiliation{%
Department of Mathematics,\\  Khalifa University of Science and Technology,\\ Main Campus, Abu Dhabi,\\ United Arab Emirates}
\author{M. Nowakowski}
\email{mnowakos@uniandes.edu.co}
\affiliation{
Departamento de Fisica,\\ Universidad de los Andes, Cra.1E
No.18A-10, Bogota, Colombia
}
\author{H. Kittaneh}
\email{Hadeel Ali Kittaneh [100045809@ku.ac.ae]}
\affiliation{%
Department of Mathematics,\\  Khalifa University of Science and Technology,\\ Main Campus, Abu Dhabi,\\ United Arab Emirates}

\date{\today}

\date{\today}

\begin{abstract}
In this work, we study the weak and strong gravitational lensing in the presence of an accelerating black hole in a universe with positive cosmological constant $\Lambda$. First of all, we derive new perturbative formulae for the event and cosmological horizons in terms of the Schwarzschild, cosmological and acceleration scales. In agreement with previous results in the literature, we find that null circular orbits for certain families of orbital cones originating from a saddle point of the effective potential are allowed and they do not exhibit any dependence on the cosmological constant. They turn out to be Jacobi unstable. We also show that it is impossible to distinguish a $C$-black hole from a $C$-black hole with $\Lambda$ if we limit us to probe only into effects associated to the Sachs optical scalars. This motivates us to analyze the weak and strong gravitational lensing when both the observer and the light ray belong to the aforementioned family of invariant cones. In particular, we derive analytical formulae for the deflection angle in the weak and strong gravitational lensing regimes.

\end{abstract}

\maketitle
\
\section{Introduction}
The $C$-metric with a positive cosmological constant $\Lambda$ is a special case of the Pleb$\acute{\mbox{a}}$nski-Demi$\acute{\mbox{a}}$nski family of metrics \cite{Griff0}. It can be obtained from the line element (23) in \cite{Griff} by setting the rotation parameter  $a=0$ and further imposing that the electric and magnetic charges vanish, that is $e=g=0$ and it describes two causally disconnected black holes of mass $M$ each accelerating in opposite direction due to the presence of a force generated by conical singularities located along the axes $\vartheta=0$ and $\vartheta=\pi$ \cite{Griff}. More precisely, in Boyer-Lindquist coordinates and in geometric units ($c=G_N=1$) the line  element is expressed as
\begin{equation}\label{metricCL}
ds^2=g_{\mu\nu}dx^\mu dx^\nu=F(r,\vartheta)\left[-f_\Lambda(r)dt^2+\frac{dr^2}{f_\Lambda(r)}+\frac{r^2}{g(\vartheta)}d\theta^2+r^2g(\vartheta)\sin^2{\vartheta}d\varphi^2\right]
\end{equation}
with
\begin{equation}\label{Ffg}
F(r,\vartheta)=(1+\alpha r\cos{\vartheta})^{-2},\quad
f_\Lambda(r)=\left(1-\frac{2M}{r}\right)\left(1-\alpha^2 r^2\right)-\frac{\Lambda}{3}r^2,\quad
g(\vartheta)=1+2\alpha M\cos{\vartheta},
\end{equation}
where $\vartheta\in (0,\pi)$, $\varphi\in(-k\pi,k\pi)$ and $r_H<r<r_h$. Note that $r_H$ denotes the event horizon, $\alpha$ is the acceleration parameter and $r_c$ is the cosmological horizon. We present a detailed analysis of the horizons and their spatial ordering in terms of the relevant physical parameters in the section. Here, it suffices to mention that while for the case of a $C$-metric the Schwarzschild horizon is smaller than the acceleration horizon whenever $0<2\alpha M<1$, it is not clear a priori if  the same condition ensures that $r_H<r_c$  in the case of a $C$-metric with positive cosmological constant. Furthermore, by adapting the tetrad (7) in \cite{Griff} to the present case the only non-zero component of the Weyl tensor is
\begin{equation}
\Psi_2=-M\left(\frac{1+\alpha r\cos{\vartheta}}{r}\right)^3.
\end{equation}
The above expression confirms that the spacetime with line element (\ref{metricCL}) is of algebraic type D and the only curvature singularity occurs at $r=0$. Hence, the horizons $r_H$ and $r_C$ are just coordinate singularities. It is interesting to observe that for $\alpha\to 0$ the metric in  (\ref{metricCL}) becomes the metric of a Schwarzschild-de Sitter black hole while in the case of $\Lambda\to 0$ but $\alpha\neq 0$ (\ref{metricCL}) correctly reproduces the line element associated to the $C$-metric as given in \cite{Maha}. This fact means that any prediction regarding the bending of light in a manifold described by (\ref{metricCL}) should reproduce the corresponding results for the Schwarzschild-de Sitter case in the limit of $\alpha\to 0$ as well as the  gravitational lensing results for the $C$-metric obtained in \cite{Maha} when we let $\Lambda\to 0$ with $\alpha$ kept constant. According to \cite{Griff}, the conical singularity on $\vartheta=0$ can be removed by making the following choice for the parameter $k$ entering in the range of the angular variable $\varphi$
\begin{equation}
k=\frac{1}{1+2\alpha M},
\end{equation}
while the conical singularity with constant deficit angle along the half-axis $\vartheta=\pi$ which is computed by means of (22) in \cite{Griff} as
\begin{equation}
\delta=\frac{8\pi\alpha M}{1+2\alpha M},
\end{equation}
can be explained in terms of a semi-infinite cosmic string pulling the black hole and/or of a strut pushing it. Similarly as for the $C$-metric, this allows us to think of (\ref{metricCL}) as of a Schwarzschild-de Sitter-like black hole experiencing an acceleration along the $\vartheta=\pi$ direction due to the presence of a force, i.e. the tension of a cosmic string. Moreover, we observe that the length of interval for the range of the coordinate $\varphi$ can be transformed to its standard value $2\pi$ with the help of the rescaling $\varphi=k\phi$ so that $\phi\in(-\pi,\pi)$. Moreover, \cite{Kof} was able to provide an interpretation of the string/strut in terms of null dust. In the rest of this paper, we will work with the line element obtained after the aforementioned rescaling is introduced, namely
\begin{equation}\label{metricFinal}
ds^2=-B_\Lambda(r,\vartheta)dt^2+A_\Lambda(r,\vartheta)dr^2+C(r,\vartheta)d\theta^2+D(r,\vartheta)d\phi^2
\end{equation}
where
\begin{equation}\label{coefficienti}
B_\Lambda(r,\vartheta)=f_\Lambda(r)F(r,\vartheta),\quad
A_\Lambda(r,\vartheta)=\frac{F(r,\vartheta)}{f_\Lambda(r)},\quad
C(r,\vartheta)=r^2\frac{F(r,\vartheta)}{g(\vartheta)},\quad
D(r,\vartheta)=k^2r^2 g(\vartheta)F(r,\vartheta)\sin^2{\vartheta}
\end{equation}
with $F$, $f_\Lambda$ and $g$ given in (\ref{Ffg}).

In the present work, we study the geodesic motion of a massive particle and the light bending in a two black hole metric with positive cosmological constant. For this end, a preliminary study of the behaviour of the null geodesics turns out to be convenient in detecting some features of strong gravity in the aforementioned spacetime. The question of new phenomena  arises if we consider a metric which in some limiting case reduces to the Schwarzschild metric (for examples see \cite{we1}). Here, the fate of the circular orbit, already appearing in the Schwarzschild metric, and issues regarding its stability deserve careful attention because they will give us useful insights on how to construct an appropriate impact parameter. While the $C$-metric has been extensively studied in the last decades, the same cannot be said for its counterpart with $\Lambda$. The independence of null geodesics on the cosmological constant was first recognized in a seminal and extremely  comprehensive paper on photon surfaces by \cite{Claudel} where a general class of static spherically symmetric space-times was considered. The case of non spherically symmetric manifolds was addressed by \cite{Gibbons}. There, among several physically relevant space-times, the case of a $C$-metric with cosmological constant was studied and the authors discovered that a non-spherically symmetric photon surface continues to exist even in that scenario. The conclusion is that, while the Schwarzschild metric exhibits a photon sphere, the same cannot be said for the $C$-metric. More precisely, \cite{Gibbons} showed that instead of a photon sphere there is a photon surface displaying at least one conical singularity.

Regarding geodesic motion in a $C$-black hole, radial time-like geodesics were analyzed by \cite{Far} whereas the study of the circular motion of massive and massless particles was undertaken by \cite{Pravda}. Moreover, \cite{Lim,Bini0} offered an exhaustive treatment of time-like and null geodesics by a mixture of analytical and numerical methods. Finally, \cite{Bini} probed into the motion of spinning particles around the direction of acceleration of the black hole. Furthermore, \cite{Maha} determined the coordinate angle of the so-called photon cone and performed a Jacobi stability analysis to show that all circular null geodesic on the photon cone are radially unstable. The shadow of a $C$-black hole was studied in \cite{Gr1,Gr2} while \cite{Frost} derived inter alia an exact solution of the light-like geodesic equation by means of Jacobi elliptic functions and determined the angular radius of the shadow. We should also mention that the analysis of the light-like geodesics and the black hole shadow for a rotating $C$-metric have been addressed in \cite{ZZ}. Moreover, \cite{other} probed into photon spheres and black hole shadows for dynamically evolving spacetimes. Finally, we refer to \cite{QNMs1,QNMs2} for the analysis of the QNMs and the stability properties for a $C$-black hole.

Regarding a $C$-black hole in an anti-de Sitter (AdS) or a de Sitter (dS) background, \cite{Griff1,Griff2,Dias1,Dias2,Krt1,Poldo,Krt2,Xu} studied in detail the geometric structure and the related properties of these spacetimes. The analysis of the circular motion of massive and massless particles in the $C$-metric with a negative cosmological constant was performed by \cite{Chamb} where the author proved that the circular null geodesics are unstable whereas \cite{Poldo} completed the study of \cite{Chamb} by considering some special characteristics associated to the null geodesics in the aforementioned metric. Recently, Lim in \cite{Lim21} classified all possible trajectories for photons in the (A)dS $C$-metric in terms of the particle angular momentum and the energy scaled in units of the Carter constant. Similarly as in \cite{Frost}, it was possible to construct exact solutions for null geodesics by means of Jacobi elliptic functions. However, the stability problem for such trajectories has not been addressed by \cite{Lim21}. We complete the results of \cite{Lim21} by concentrating on the interplay of different scales in the potentially observable or physical relevant case.  To make a comparison with known results of the Schwarzschild- de Sitter metric more accessible, we work in Boyer-Lindquist coordinates.

Concerning gravitational lensing, detailed studies on light bending in the weak and strong regimes were pioneered by
the George Ellis lensing group. Some of their exciting results which have relevance to the present work are  \cite{Virb,Virb1,Virb2,Virb3,Virb4}. A nice and thorough review article on this subject has been written by \cite{Perl}. Finally, \cite{Gibbons01} exploited a novel geometrical approach based on the Gauss-Bonnet theorem applied to the optical metric of the gravitational lens in order to derive weak leansing formulae for spherically symmetric metrics generated by certain static, perfect non-relativistic fluids. Regarding the gravitational lensing \cite{Frost} derived a lens equation and showed that the lens results of \cite{Ifti} for the rotating $C$-metric with NUT parameter does not contain as a special case the $C$-metric (where both the acceleration and NUT parameters are set equal to zero). Moreover, \cite{Maha} studied the strong and weak lensing for null rays on the photon cone. To the best of our knowledge, we could not identify any paper studying the bending of light and analyzing the (in)stability problem of null circular orbits for the $C$-metric with positive cosmological constant. We hope to fill this gap with the present work.

The remainder of the paper is structured as follows. In Section II, we analyze the horizon structure of the dS $C$-metric in terms of certain orderings among the Schwarzschild, the cosmological and the acceleration scales. In order to understand which scale orderings are physically relevant, we consider three typical black hole representatives: ultramassive, massive and light. In Section III, we derive the effective potential for massive and massless particles and we show that the null-orbits have the same radius and and take place on the same family of invariant cones as in the $C$-metric, i.e. they do not depend on $\Lambda$. This result is in agreement with \cite{Gibbons}. Moreover, we find  that the circular orbit is due to a saddle point in the effective potential,  which requires an additional effort to probe into the associated stability problem. This is addressed in Section IV where we perform the Jacobi (in)stability analysis of the circular orbits. In Section V,  since the Sachs optical scalars cannot be used to optically distinguish between $C$- and a dS $C$-black holes, we study the the gravitational lensing in the weak and strong regimes. More precisely, the corresponding deflection angles are analytically computed when the light propagation occurs on a certain family of invariant cones, and  their dependence on the observer position is shown. Our formulae correctly reproduce the corresponding ones in the $C$-metric case in the limit of vanishing $\Lambda$ and indicate that the deflection angles may depend on the cosmological constant if the position of the observer is close to the cosmological horizon.

\section{Analysis of the horizons}\label{analisi} 
The structure of the horizons for the metric associated to the line element (\ref{metricCL}) can be unraveled by analyzing the roots of the equation $f_\Lambda(r)=0$ with $f_\Lambda$ given as in (\ref{Ffg}). To this purpose, it is convenient to introduce the Schwarzschild, the cosmological and the acceleration length scales defined as $r_s=2M$, $r_\Lambda=\sqrt{3/\Lambda}$ and $r_a=1/\alpha$, respectively. The appearance of several scales in the metric makes a precise study of the horizons a worthwhile undertaking since a priori it is not clear what structure of the horizons will emerge. Apart from that, we recall a curious fact from the Schwarzschild de Sitter metric with two horizons, one dominated by $r_s$ and the second one (the cosmological horizon) by $r_{\Lambda}$.  The Boyer-Lindquist coordinates which one uses to study this metric are valid within these two horizons where we locate ourselves and the rest of the universe. An observer outside $r_{\Lambda}$ might even claim that we are living inside a black hole. It is interesting to reconsider the unique position for the case of the C-metric as more scales enter the calculation. In terms of these scales the equation $f_\Lambda(r)=0$ gives rise to the following cubic equation
\begin{equation}\label{kub}
P(r)=r^3-\rho r^2-\sigma r+\tau=0
\end{equation}
with
\begin{equation}
\rho=\frac{r_{s}r_{\Lambda}^2}{r_{\Lambda}^2+r_a^2}, \quad
\sigma=\frac{r_{a}^2r_{\Lambda}^2}{r_{\Lambda}^2+r_a^2},\quad
\tau= \frac{r_{s}r_{a}^2r_{\Lambda}^2}{r_{\Lambda}^2+r_a^2}.
\end{equation}
First of all, we observe that the extrema of the cubic in (\ref{kub}) are located at
\begin{equation}
r_\pm=\frac{r_s r_\Lambda^2}{3(r_{\Lambda}^2+r_a^2)}\left[1\pm\sqrt{1+\Delta}\right],\quad
\Delta=3\frac{r_a^2}{r_s^2}\left(1+\frac{r_a^2}{r_\Lambda^2}\right)>0
\end{equation}
with $r_{-}<0$, $r_{+}>0$ and moreover,
\begin{equation}\label{zanimivo}
P(r_{-})=\frac{r_\Lambda^2}{27(r_\Lambda^2+r_a^2)^3}\left\{
2r_s^3 r_\Lambda^4\left[\left(1+\Delta\right)^{3/2}-1\right]+18r_s r_a^2 r_\Lambda^4 +45r_s r_\Lambda^2 r_a^4+27r_s r_a^6
\right\}
\end{equation}
Since $P(r)\to-\infty$ as $r\to-\infty$, we conclude that $r_{-}$ is a maximum. The fact that $P(r_{-})>0$ for any positive value of the scales, as it can be immediately seen from (\ref{zanimivo}), together with the observation that $P(0)>0$, allows us to conclude that the cubic (\ref{kub}) admits always a negative root, here denoted by $r_{<}$ while the other two zeroes may be positive and distinct, having algebraic multiplicity $2$ or being complex conjugate of each other.  However,the scenario where the two positive roots coincide, i.e. the discriminant of (\ref{kub}) vanishes, is not physically relevant because as the discriminant tends to zero, the region between the two positive roots gets smaller and smaller and it is impossible to define a static observer in most of the spacetime. Furthermore, the case of two complex conjugate roots corresponds to the presence of a naked singularity at $r=0$. Since the cubic (\ref{kub}) depends on the three scales $r_s$, $r_a$, and $r_\Lambda$, it is imperative to understand which orderings among these scales are physically relevant. To this purpose, we consider three typical black hole representatives: ultramassive BHs such as TON618 in Canes Venatici \cite{TON} whose Schwarzschild radius is $32$ times the distance from Pluto to the Sun (see Table~\ref{tableNull}), massive BHs like Sagittarius A$^{*}$ at the galactic centre of the Milky Way \cite{SAG} whose event horizon is approximately $18$ times the sun radius and light BHs such as 
GW170817 in the shell elliptical galaxy NGC 4993 \cite{ABB} with an event horizon diameter of $18$ km.
\begin{table}[ht]
\caption{Typical values of the scales and acceleration parameter for different black hole scenarios. Here, $M_\odot=1.989\cdot 10^{30}$ Kg and $r_\odot=6.957\cdot 10^8$ m denote the solar mass and the sun radius, respectively. The value for the cosmological constant is taken to be $\Lambda\approx 10^{-52}$ m$^{-2}$ as in \cite{Carmeli} while the values of the ratios $M/M_\odot$ are as given in \cite{TON,SAG,ABB}. The fifth column represents the allowed ranges for the acceleration parameter $\alpha_c$ in the case the aforementioned black holes are modelled in terms of the $C$-metric for which it is necessary to consider the constraint $2M\alpha_c<1$. From the last column where $\alpha_\Lambda=c^2\sqrt{\Lambda/3}$ we see that the case $r_s\ll r_a=r_\Lambda$ can only be relevant to light black holes such as GW170817.}
\begin{center}
\begin{tabular}{ | l | l | l | l|l|l|l|}
\hline
$\mbox{BH name}$             & $M/M_\odot$ & $r_s$ (m)     & $r_s/r_\Lambda$ &  $\alpha_c$ (m/s$^2$)   & $\alpha_s$ (m/s$^2$) &$2G_N M \alpha_\Lambda/c^2$  \\ \hline
\mbox{TON618}                & $6.6\cdot 10^{10}$  &  $1.9\cdot 10^{14}$ & $1.1\cdot 10^{-12}$ & $<5.1\cdot 10^{-15}$& $5.1\cdot 10^{-15}$& $10^5$ \\ \hline
\mbox{Sagittarius A$^{*}$}   & $4.3\cdot 10^{6}$   &  $1.3\cdot 10^{10}$ & $7.3\cdot 10^{-17}$ & $<7.8\cdot 10^{-10}$& $7.8\cdot 10^{-10}$& $7$ \\ \hline
\mbox{GW170817}              & $2.74$              &  $8.1\cdot 10^3$    & $4.7\cdot 10^{-23}$ & $<1.2\cdot 10^{-4}$& $1.2\cdot 10^{-4}$ &$10^{-6}$\\ \hline
\end{tabular}
\label{tableNull}
\end{center}
\end{table}
From Table~\ref{tableNull}, we immediately observe that we can always assume $r_s\ll r_\Lambda$. Regarding the acceleration parameter $\alpha$, it is important to observe that for $C$-black holes with cosmological constant the only constraint we need to impose on $\alpha$ is that $\alpha>0$. The situation is dramatically different in the case of the $C$-metric where the black hole mass and the acceleration parameter must satisfy the condition $2M\alpha<1$ which is equivalent to require that $r_s<r_a$. Let us discuss and interpret the roots of (\ref{kub}) for the following cases
\begin{enumerate}
\item
$r_s=r_a\ll r_\Lambda$: in this scenario, given $M$ the acceleration parameter of the black hole in SI units is
\begin{equation}
\alpha_s=\frac{c^2}{2G_N M}
\end{equation}
See Table~\ref{tableNull} for typical values for $\alpha_s$. Note that this case has no corresponding physical counterpart for a $C$-metric because in the limit of $\Lambda\to 0$ we would have a $C$-BH such that the event and acceleration horizons coincide. Setting $r_s=r_a$ in (\ref{kub}) and introducing the small parameter $\epsilon=r_s/r_\Lambda$, the discriminant of the reduced cubic is
\begin{equation}
D_1=\frac{\epsilon^2 r_s^6(27\epsilon^2+32)}{108(1+\epsilon^2)^4},
\end{equation}
which is clearly positive. This observation together with the remark below equation (\ref{zanimivo}) allows us to conclude that there is one negative root and two complex conjugate roots. Hence, this is the case of a naked singularity at $r=0$ and the coordinate $r$ can be extended up to space-like infinity. In comparison a naked singularity is not possible in the Schwarzschild-de Sitter metric.
\item
$r_s<r_a\ll r_\Lambda$: if we rewrite (\ref{kub}) as
\begin{equation}
-\frac{\Lambda}{3}r^3+(r-r_s)\left(1-\frac{r^2}{r_a^2}\right)=0,
\end{equation}
we see that $\Lambda$ is the small parameter and a straightforward application of perturbation methods for algebraic equations shows that the event horizon $r_H$ and the cosmological horizon $r_h$ are represented by the following expansions
\begin{eqnarray}
r_H&=&r_s+\frac{r_s^3 r_a^2}{3(r_a^2-r_s^2)}\Lambda+\frac{r_s^5(3r_a^2-r_s^2)}{(r_a^2-r_s^2)^3}\Lambda^2+\mathcal{O}(\Lambda^3),\\
r_h&=&r_a-\frac{r_a^4}{6(r_a-r_s)}\Lambda+\frac{r_a^7(3r_a-5r_s)}{72(r_a-r_s)^3}\Lambda^2+\mathcal{O}(\Lambda^2).
\end{eqnarray}
From the point of view of the horizon structure this is an interesting case.  Even if $\Lambda$ appears in the corrections, the horizon associated with it disappears and in its place, we encounter $r_a$ which we could rightly call the acceleration horizon.  As a consequence, we locate our position within the acceleration horizon. Note that the same expansion holds also for $r_a<r_s\ll r_\Lambda$.
\item
$r_s\ll r_a=r_\Lambda$: in this regime the acceleration parameter is completely determined by the cosmological constant and is given in SI units as
\begin{equation}
\alpha_\Lambda=c^2\sqrt{\frac{\Lambda}{3}}\approx 5.2\cdot 10^{-10}~\mbox{m/s$^2$}.
\end{equation}
Since $r_s\ll r_a$, it must be $2M\alpha_\Lambda\ll 1$. As we can see from the last column in Table~\ref{tableNull}, such a condition is violated by ultramassive and massive black holes. Hence, the present case may be relevant for light black holes such as GW170817. By means of the rescaling $\mathfrak{r}=r/r_\Lambda$ and the introduction of the same small parameter $\epsilon$ already defined  in 1. we can rewrite (\ref{kub}) as
\begin{equation}\label{pertur}
\frac{\epsilon}{2}(1-\mathfrak{r}^2)+\mathfrak{r}\left(\mathfrak{r}^2-\frac{1}{2}\right)=0.
\end{equation}
The discriminant of the associated reduced cubic is 
\begin{equation}\label{D3}
D_3=-\frac{\epsilon^4}{432}+\frac{71}{1728}\epsilon^2-\frac{1}{216}.
\end{equation}
From Fig.~\ref{f02} we observe that also in this case the discriminant may become positive. More precisely, we have three distinct real roots if $0\leq\epsilon<\epsilon_0=\sqrt{142-34\sqrt{17}}/4$ while the naked singularity case occurs when $\epsilon>\epsilon_0$. Applying again perturbation methods to find expansions for the positive roots of (\ref{pertur}) yields
\begin{figure}[ht]\label{hic_02}
\includegraphics[scale=0.35]{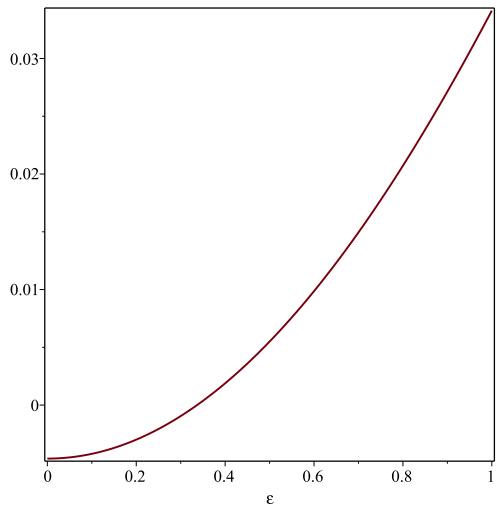}
\caption{\label{f02}
Plot of the discriminant (\ref{D3}) for $0\leq\epsilon\leq 1$.}
\end{figure}
\begin{eqnarray}
r_H&=&r_s+\frac{r_s^3}{r_\Lambda^2}+4\frac{r_s^5}{r_\Lambda^4}+\mathcal{O}\left(\frac{r_s^7}{r_\Lambda^6}\right),\label{radice1}\\
r_h&=&\frac{r_\Lambda}{\sqrt{2}}-\frac{r_s}{4}+\frac{7}{16\sqrt{2}}\frac{r_s^2}{r_\Lambda}-\frac{1}{2}\frac{r_s^3}{r_\Lambda^2}-\frac{689}{512\sqrt{2}}\frac{r_s^4}{r_\Lambda^3}+\mathcal{O}\left(\frac{r_s^5}{r_\Lambda^4}\right).\label{radice2}
\end{eqnarray}
This case resembles indeed the Schwarzschild-de Sitter order of horizons.
\end{enumerate}
We conclude this section by observing that in general equation (\ref{kub}) can be transformed into the reduced third order polynomial equation 
\begin{equation}
Y^3+pY+q=0,\quad
p=-\frac{r_\Lambda^2(3r^2_\Lambda r_a^2+r_\Lambda^2 r_s^2+3r_a^4)}{(r_\Lambda^2+r_a^2)^2},\quad
q=\frac{r_s r^2_\Lambda\left[2r^4_\Lambda(9r_a^2-r_s^2)+45r^2_\Lambda r_a^4+27 r_a^6\right]}{27(r_\Lambda^2+r_a^2)^3}
\end{equation}
by means of the variable transformation $Y=r+\rho/3$. According to \cite{Bron} the associated discriminant is 
\begin{equation}
D=\left(\frac{p}{3}\right)^3+\left(\frac{q}{2}\right)
\end{equation}
and we have the following classification
\begin{enumerate}
\item
three distinct real roots for $D<0$;
\item
two real roots where one root has algebraic multiplicity two whenever $D=0$;
\item
one real and two complex conjugate roots for $D>0$.
\end{enumerate}
Since the first case is physically relevant, we will stick to the condition $D<0$. Then, if we introduce the additional parameter $R=\sqrt{|p|/3}$ and the auxiliary angle $\omega$ defined as $\cos{\omega}=q/(2R^3)$, then the roots are parametrized with the help of trigonometric functions and their inverses in the following form
\begin{equation}
r_1=-\frac{\rho}{3}-2R\cos{\left(\frac{\omega}{3}\right)},\quad
r_2=-\frac{\rho}{3}+2R\cos{\left(\frac{\pi}{3}-\frac{\omega}{3}\right)},\quad
r_3=-\frac{\rho}{3}+2R\cos{\left(\frac{\pi}{3}+\frac{\omega}{3}\right)}.
\end{equation}
We observe that in addition to the inequality $D<0$, there is the additional constraint that $\cos{\omega}\leq 1$. These two constraints are not satisfied for any value of the scales entering in our problem as it can be seen in Figure~\ref{fff}. However, it can be seen that the condition $D<0$ ensures that $\cos{\omega}< 1$. 
\begin{figure}[ht]\label{hic_hic}
\includegraphics[scale=0.35]{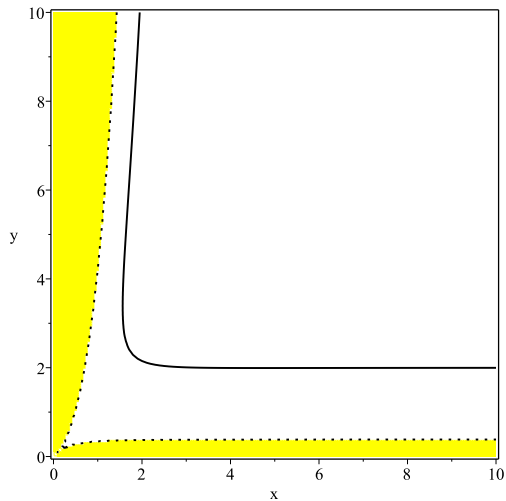}
\caption{\label{fff}
The yellow region represents the region in the parameter space $(x,y)$ with $x=r_a/r_\Lambda$ and $y=r_s/r_\Lambda$ where the constraints $D<0$ and $\cos{\omega}\leq 1$ are simultaneously satisfied. The solid black line is the curve along which $\cos{\omega}=1$.}
\end{figure}

\section{Geodesic equations and effective potential}
\begin{figure}[ht]\label{hierhier}
\includegraphics[scale=0.35]{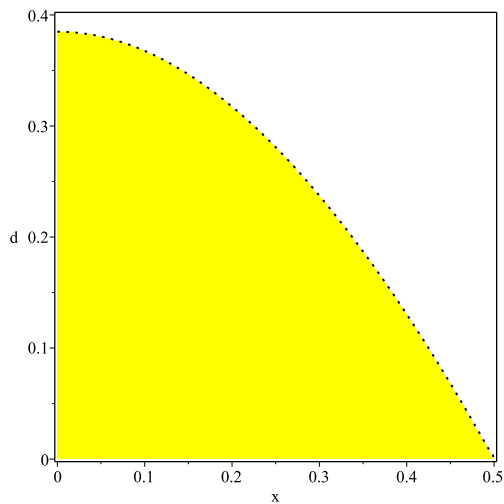}
\caption{\label{fig0}
The yellow region represents those points $(x,d)$ for which the function $S(x)+K(d,x)$ appearing in the Hessian (\ref{Hessian}) is positive.}
\end{figure}

When we turn our attention to the study of the geodesic motion, in addition to $r_s$, $r_{\Lambda}$ and $r_a$, a new scale $\ell$ associated to the angular momentum of the particle enters the scene. To probe into the interplay of these scales, the method of the effective potential seems most adequate.  One might expect that the many new scales as compared to the Schwarzschild-de Sitter metric will result into a richer structure of critical points. It will turn out that this is partly  true, but only if we pay careful attention to the emergence of saddle points. To study the motion of a particle in the gravitational field described by (\ref{metricFinal}), we need to analyze the geodesic equations \cite{FL}
\begin{equation}\label{geo}
\frac{d^2x^\eta}{d\lambda^2}+ \Gamma^\eta{}_{\mu\nu} \frac{dx^\mu}{d\lambda} \frac{dx^\nu}{d\lambda}=0,\quad
\frac{1}{2}g^{\eta\tau}\left(\partial_\mu g_{\tau\nu}+\partial_\nu g_{\tau\mu}-\partial_\tau g_{\mu\nu}\right)
\end{equation}
subject to the constraint
\begin{equation}\label{constraint}
 g_{\mu\nu}\frac{dx^\mu}{d\lambda} \frac{dx^\nu}{d\lambda}=-\epsilon,
\end{equation}
with $\epsilon = 0$ and $\epsilon = 1$ for light-like and time-like particles, respectively. The system of coupled ODEs associated to (\ref{geo}) can be immediately obtained from equations (8)-(11) in \cite{Maha} by replacing $A$, $B$ and $f$ therein with the functions $A_\Lambda$, $B_\Lambda$ and $f_\Lambda$ defined in (\ref{coefficienti}) and noticing that the functions $C$ and $D$ remain the same. In view of this observation, one can proceed as in \cite{Maha} and conclude that the dynamics is governed by the following coupled system of ODE 
\begin{eqnarray}
\frac{d^2 r}{d\lambda^2}&=&-\frac{\partial_r A_\Lambda}{2A_\Lambda}\left(\frac{dr}{d\lambda}\right)^2-\frac{\partial_\vartheta A_\Lambda}{A_\Lambda}\frac{dr}{d\lambda}\frac{d\vartheta}{d\lambda}+\frac{\partial_r C}{2A_\Lambda}\left(\frac{d\vartheta}{d\lambda}\right)^2
-\frac{\mathcal{E}^2}{2}\frac{\partial_r B_\Lambda}{A_\Lambda B_\Lambda^2}+\frac{\ell^2}{2}\frac{\partial_r D}{A_\Lambda D^2},\label{g7}\\
\frac{d^2\vartheta}{d\lambda^2}&=&-\frac{\partial_\vartheta C}{2C}\left(\frac{d\vartheta}{d\lambda}\right)^2-\frac{\partial_r C}{C}\frac{dr}{d\lambda}\frac{d\vartheta}{d\lambda}+\frac{\partial_\vartheta A_\Lambda}{2C}\left(\frac{dr}{d\lambda}\right)^2-\frac{\mathcal{E}^2}{2}\frac{\partial_\vartheta B_\Lambda}{CB^2_\Lambda}+\frac{\ell^2}{2}\frac{\partial_\vartheta D}{CD^2},\label{g8}
\end{eqnarray}
where $\mathcal{E}$ and $\ell$ are the energy per unit mass and the angular momentum per unit mass of the particle, respectively. Moreover, the constraint equation (\ref{constraint}) can be cast into the form
\begin{equation}\label{cos}
\frac{F^2}{2}\left[\left(\frac{dr}{d\lambda}\right)^2+\frac{r^2 f_\Lambda}{g}\left(\frac{d\vartheta}{d\lambda}\right)^2\right]+U_{eff}=E,
\end{equation}
where $E=\mathcal{E}^2/2$ and the effective potential is given by
\begin{equation}\label{pott}
U_{eff}(r,\vartheta)=\frac{B_\Lambda}{2}\left(\epsilon +\frac{\ell^2}{D}\right).
\end{equation}
At this step, it is gratifying to observe that in the limit of vanishing $\alpha$ and $\Lambda$ equation (\ref{cos}) reproduces correctly equation (25.26) in \cite{FL} for the Schwarzschild case. Moreover, the  functions $A_\Lambda$, $B_\Lambda$ and $C$ are non negative for $r_H\leq r\leq r_h$ and therefore, $E-U_{eff} \ge 0$ as in classical mechanics. Finally, the equality, $E=U_{eff}$, corresponds to a circular orbit and a critical point of the effective potential. Since in the present work we are interested in the study of the light bending, we recall that in the case of null geodesics $\epsilon=0$ and hence, the effective potential simplifies as follows
\begin{equation}\label{Uefflight}
\mathfrak{V}(r,\vartheta)=\frac{\ell^2 B_\Lambda}{2D}.
\end{equation}
To study the null circular orbits for the potential (\ref{Uefflight}), we need to find its critical points. Imposing that $\partial_r\mathfrak{B}=0=\partial_\vartheta\mathfrak{B}$ leads to the following equations
\begin{equation}
\alpha^2 M r^2+r-3M=0,\quad
3\alpha M\cos^2{\vartheta}+\cos{\vartheta}-\alpha M=0.
\end{equation}
At this point a comment is in order. First of all, the above equations does not contain $\Lambda$. This is surprising because the spacetimes described by (\ref{metricFinal}) and the $C$-metric are not conformally related. The same phenomenon occurs when we study the null circular orbits for the Schwarzschild and Schwarzschild-de Sitter black holes, i.e. in both cases the corresponding photon spheres are characterized by a typical radius which is $\Lambda$-independent \cite{Claudel}. Hence, we can conclude as in \cite{Maha} that null geodesics admit circular orbits with radius 
\begin{equation}
r_c=\frac{6M}{1+\sqrt{1+12\alpha^2 M^2}}
\end{equation}
only for a certain family of orbital cones with half opening angle given by
\begin{equation}
    \theta_c= \arccos{\left(\frac{2\alpha M }{1+\sqrt{1+12\alpha^2M^2}}\right)}.
\end{equation}
Note that due to the fact that $\alpha M\in(0,1/2)$ and $\vartheta_c$ is a monotonically decreasing function in the variable $\alpha M$, it follows that $\vartheta_c$  cannot take every value from $0$ to $\pi$. More precisely, it can only vary on the interval $(\vartheta_{c,min},\pi/2)$ with $\vartheta_{c,min}=\arccos{(1/3)}\approx 70.52^{\circ}$. To classify the critical point of our effective potential, we compute the determinant of the Hessian matrix associated with the effective potential (\ref{Uefflight}) at the critical point $(r_c, \theta_c)$. We find that the determinant $\Delta$ of the Hessian matrix is
\begin{equation}\label{Hessian}
\Delta(r_c,\vartheta_c)=-\frac{\ell^4}{139968 M^6\kappa^4}\frac{(1+\tau)^9}{(1+\tau+4x^2)^3}\frac{S(x)+K(d,x)}{T(x)}
\end{equation}
with $x:=\alpha M$, $d=r_s/r_\Lambda$, $\tau=\sqrt{1+12x^2}$. The functions $S$ and $T$ are the same as those computed in \cite{Maha}, namely 
\begin{eqnarray}
S(x)&:=&(1728\tau+8640)x^{10}+(1008\tau-720)x^8-(492\tau+828)x^6+(25-35\tau)x^4+(13\tau+19)x^2+1+\tau,\label{S}\\
T(x)&:=&32 x^8+(32\tau+176)x^6+(48\tau+114)x^4+(14\tau+20)x^2+1+\tau,
\end{eqnarray}
while the new contribution due to the cosmological constant is encoded in the function $K(d,x)$ which is given by
\begin{equation}
K(d,x)=-d^2\left[1944 x^8+(1296\tau+4860)x^6+\left(918\tau+\frac{3375}{2}\right)x^4+\left(\frac{297}{2}\tau+1892\right)x^2+\frac{27}{4}(\tau+1)\right].
\end{equation}
Note that in the limit of $d\to 0$ equation (\ref{Hessian}) reproduces correctly (31) in \cite{Maha}. From the analysis performed in \cite{Maha} we already know that the function $T$ is always positive for $x\in(0,1/2)$. This signalizes that the sign of (\ref{Hessian}) is controlled by the term $S(x)+K(d,x)$ which is positive for $x$ in the interval $(0,1/2)$ and 
\begin{equation}
d<\mathfrak{f}(x),\quad
\mathfrak{f}(x)=2\sqrt{3}(1-2x)\sqrt{\frac{(9\tau+45)x^4+(9\tau+15)x^2+\tau+1}{24x^6+(16\tau+58)x^4+(10\tau+16)x^2+\tau+1}},
\end{equation}
where $\mathfrak{f}$ is the function representing the dotted boundary of the yellow region in Fig.~\ref{fig0}. Since black holes of astrophysical interest are characterized by $r_s\ll r_\Lambda$, this implies that $d\ll 1$. The yellow part in Fig.~\ref{fig0} represents the region in the space of the parameters $x$ and $d$ where the function $S(x)+K(d,x)$ is positive. This is clearly the case for $x\in(0,1/2)$ and $d\ll 1$. Hence, we conclude that the critical point $(r_c,\vartheta_c)$ of the effective potential is a saddle point. 

\subsection{Geodesic motion for massive particles}
\begin{figure}[ht]
\centering
\includegraphics[width=0.3\textwidth]{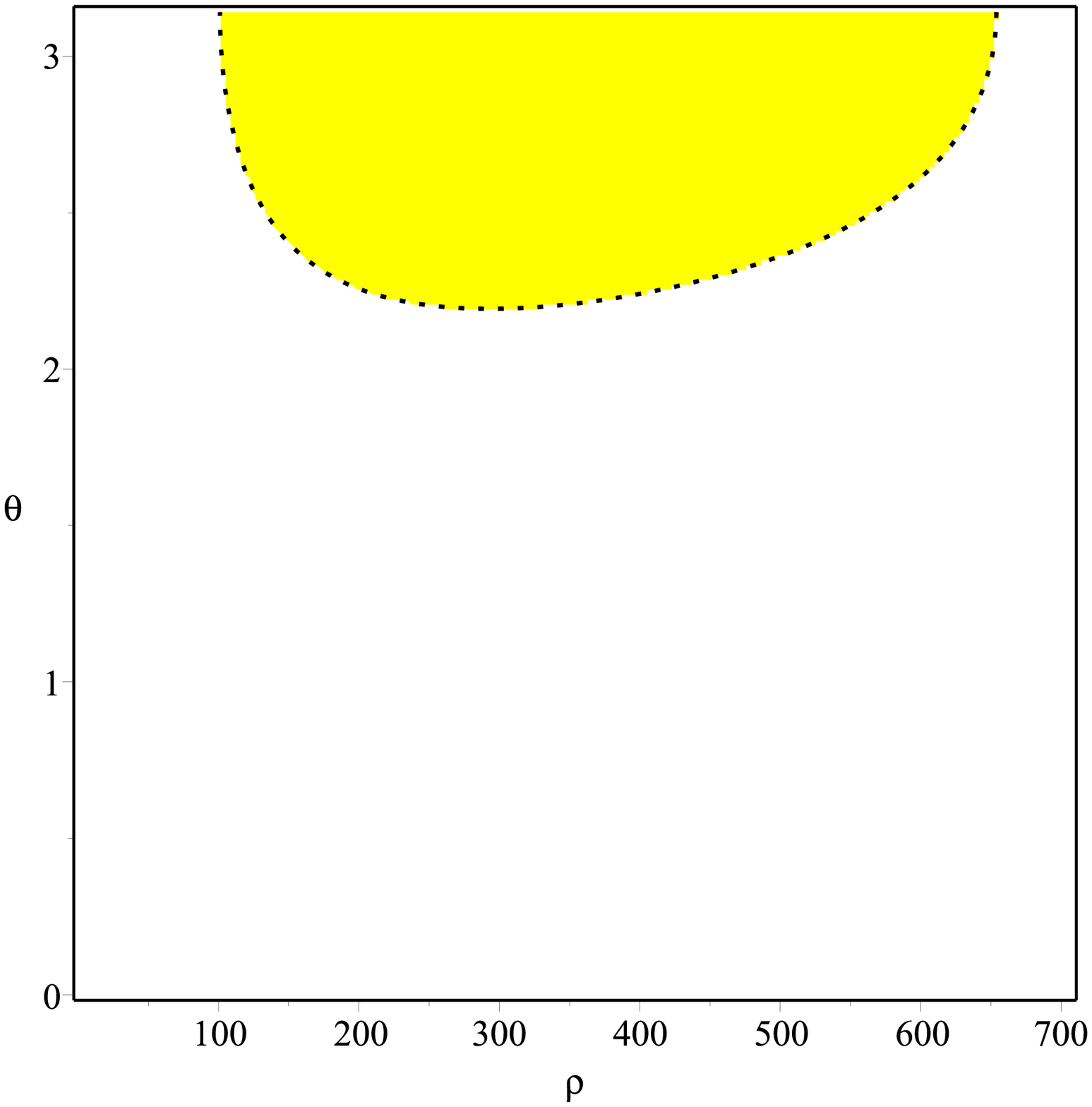}
\qquad
\includegraphics[width=0.3\textwidth]{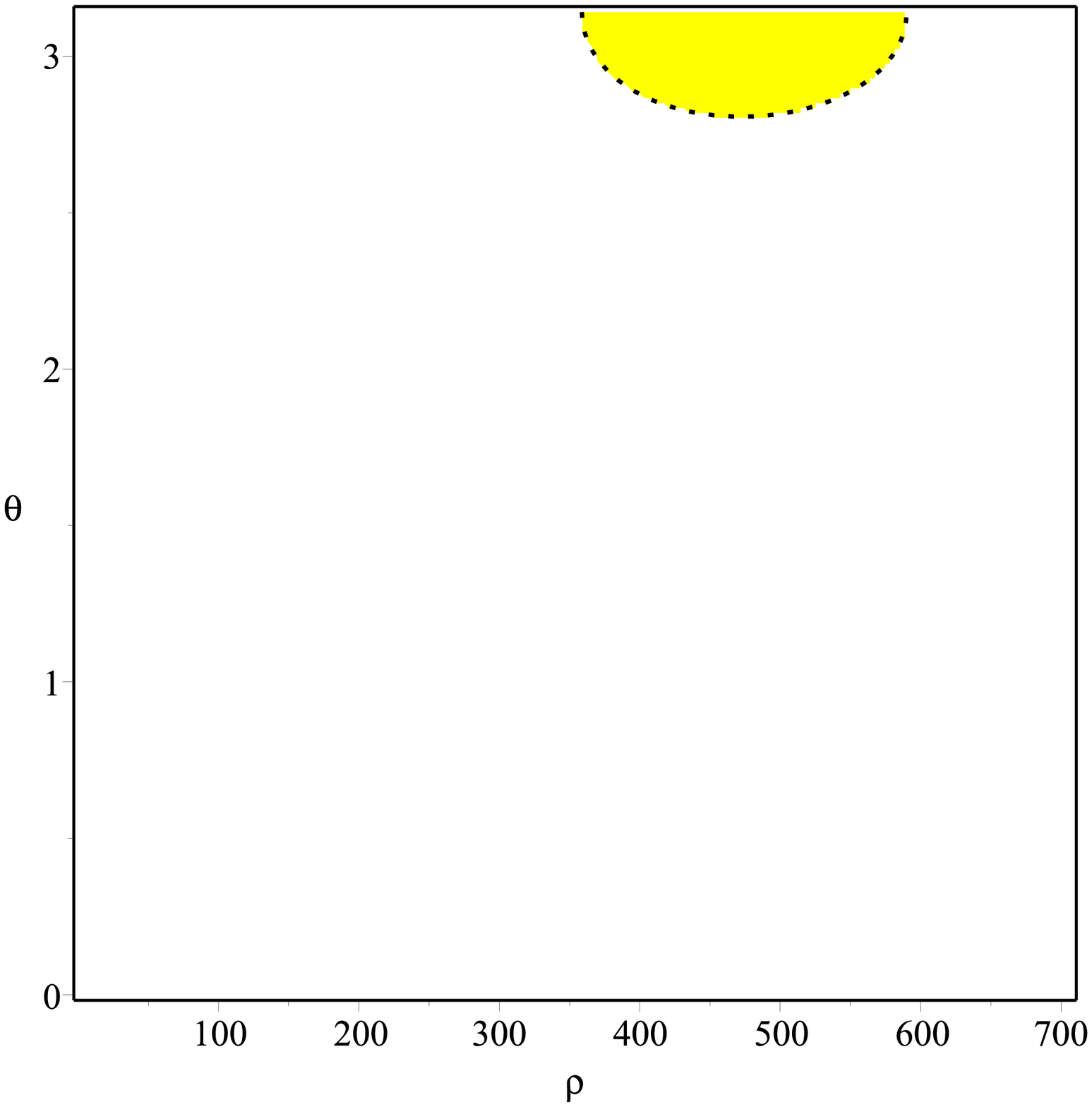}
\includegraphics[width=0.3
\textwidth]{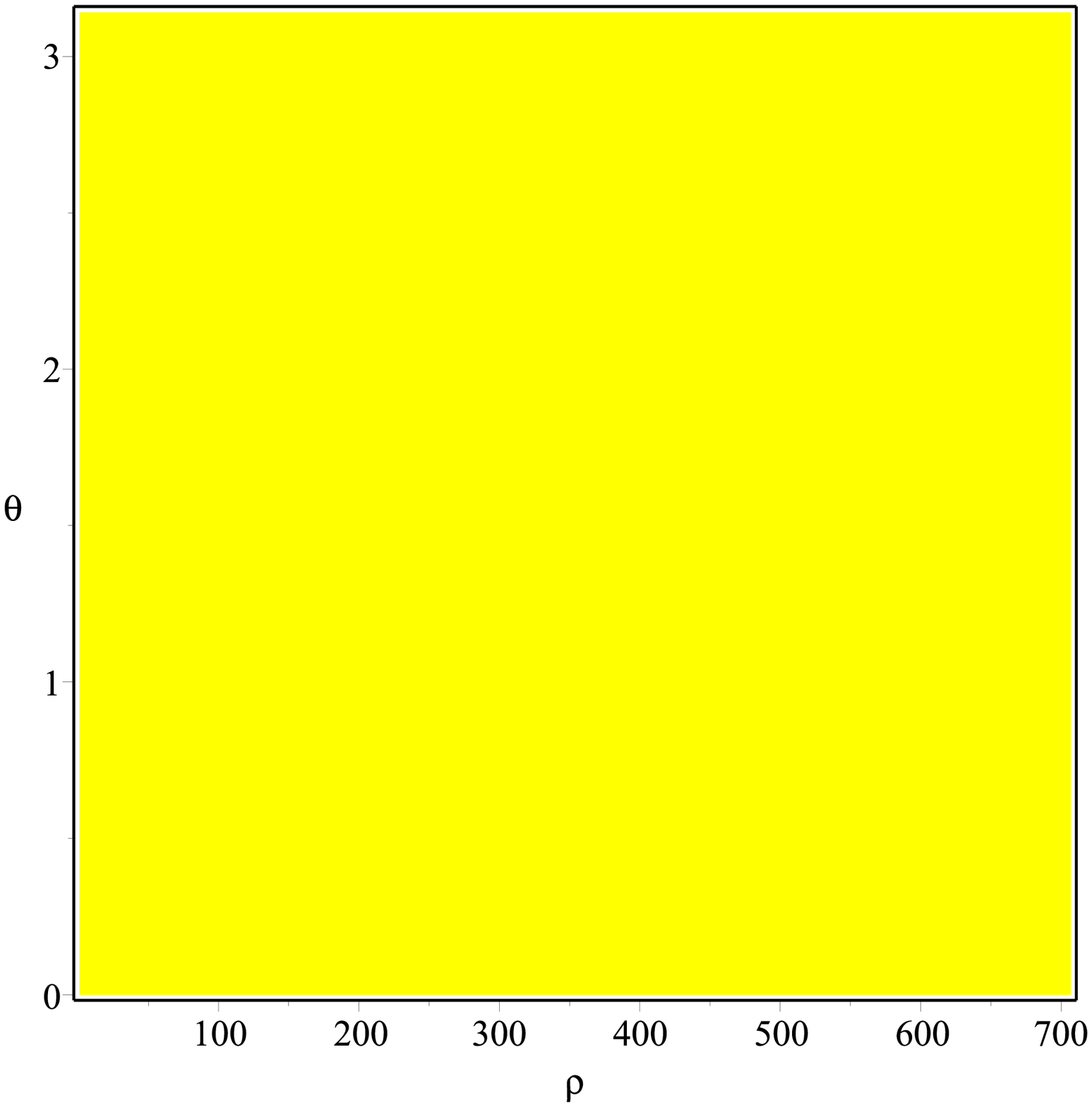}
\caption{Typical shapes of the region where the inequality $E-\mathcal{V}_{eff}>0$ is satisfied for time-like particles with $E=0.1$ (left), $E=0.49$ (middle) and $E=0.497$ (right) when $x=10^{-3}$. The event and cosmological horizons are approximately located at $\rho_H= 1+\mathcal{O}(x^2)$ and $\rho_h=\frac{1}{\sqrt{2}x}-\frac{1}{4}+\frac{7}{16\sqrt{2}}x+\mathcal{O}(x^2)\approx 706.86$.}
\label{globfigg}
\end{figure}

We study the motion of test particles for the $C$-metric with positive cosmological constant. Since there are involved three physical scales, we may expect that they may combine in such a way to lead to new results. We recall that the equation of motion for a massive particle with proper time $\tau$ in the aforementioned metric is given by (\ref{cos}) with $\lambda$ replaced by $\tau$ while the effective potential is represented by (\ref{pott}) with $\epsilon=1$. In the case $\ell=0$, the effective potential reads
\begin{equation}\label{kkk}
\mathcal{V}_{eff}(r,\vartheta)=\frac{1-\frac{r_s}{r}+\frac{r_s}{r_a^2}r-\left(\frac{1}{r^2_a}+\frac{1}{r_\Lambda^2}\right)r^2}{2\left(1+\frac{r}{r_a}\cos{\vartheta}\right)^2}.
\end{equation}
By means of the rescaling $\rho=r/r_s$, $x=r_s/r_\Lambda$ and in the regime $r_s\ll r_a=r_\Lambda$ we can cast (\ref{kkk}) into the form 
\begin{equation}\label{effkk}
\mathcal{V}_{eff}(\rho,\vartheta)=\frac{1-\frac{1}{\rho}+x^2(\rho-2\rho^2)}{2\left(1+x\rho\cos{\vartheta}\right)^2}.
\end{equation}
Concerning the behaviour of the effective potential at the cosmological horizon, we observe that at the quadratic order in the small parameter $x$ 
\begin{equation} \label{large}
\mathcal{U}_{eff}(\rho_h,\vartheta)=-\frac{7}{4(\cos{\vartheta}+\sqrt{2})}x^2+\mathcal{O}(x^3)
\end{equation}
and hence, $\mathcal{U}$ is always negative there for any $\vartheta\in[0,\pi]$. Regarding the motion of a time-like particle, it follows from (\ref{cos}) that the particle dynamics is constrained to those regions where the reality condition
\begin{equation}
E-\mathcal{V}_{eff}>0.
\end{equation}
is satisfied. In the case $r_s\ll r_a=r_\Lambda$ displayed in Fig.~\ref{globfigg}, we see that depending on the value of the parameter $E$, the geodesics can reside in different regions. For example, if $0.1<E<0.49$, the particle neither falls into the event horizon nor into the cosmological horizon. More precisely, it stays inside the yellow compact region displayed in the first two panels of Fig.~\ref{globfigg}. Such a region becomes smaller as $E$ increases. When $E$ crosses a critical value $E_{crit}\in(0.49,0.497)$, the particle will fall into the event or cosmological horizon.
Regarding the critical points of the effective potential (\ref{effkk}), the condition $\partial_\vartheta\mathcal{V}_{eff}=0$ is satisfied whenever $\vartheta=0$ or $\vartheta=\pi$. In the case $\ell=0$, it is not difficult to check that the geodesic equations (\ref{g7}) and (\ref{g8}) stay finite at the axes $\vartheta=0$ or $\vartheta=\pi$. This signalizes that mathematically speaking, we can probe into geodesics going  through the poles. As it was already noticed by \cite{Lim}, such geodesics are not physically possible because the particles moving along these trajectories would undergo a collision with the cosmic string/strut causing the black hole to accelerate. This problem can be circumvented if we imagine these time-like geodesics to be arbitrarily close to the axis $\vartheta=0$ or $\vartheta=\pi$, while keeping the geodesic equations at $\vartheta=0$ or $\vartheta=\pi$ as an approximation. In the following, we focus on the case $r_s\ll r_a=r_\Lambda$ which has not been covered by \cite{Lim21}. 
\subsubsection{Time-like radial geodesics along $\vartheta=0$}
If we impose that $\partial_r\mathcal{V}_{eff}=0$ along the north pole, we end up with the cubic equation 
\begin{equation}\label{cubo1}
x^2(x+4)\rho^3+x(2-x)\rho^2-3x\rho-1=0.
\end{equation}
Descartes' rule of signs implies that that there is only one positive root, here denoted by $\rho_{crit}$ because the polynomial (\ref{cubo1}) exhibits only one sign change due to the fact that $x\ll 1$ ensures that the term $2-x$ is positive. On the other hand, 
\begin{equation}
\left.\frac{d^2\mathcal{V}_{eff}(\rho,0)}{d\rho^2}\right|_{\rho=\rho_{crit}}=-\frac{1}{\rho^3_{crit}}+x^2+\mathcal{O}\left(x^3\right)
\end{equation}
is negative, and therefore, the equilibrium point $\rho_{crit}$ is unstable. This also signalizes that $\rho_{crit}$ is a maximum for the effective potential.  This implies that the associated geodesic is unstable and under any small perturbation, the particle will either cross the event horizon of the black hole or approach the acceleration horizon. The same behaviour occurs in the case of a vanishing cosmological constant. The latter scenario was studied in \cite{Lim}. For typical values of $\rho_{crit}$ we refer to Table~\ref{tablerhocrit}.
\begin{table}[ht]
\caption{Typical values for the location of the maximum ($\rho_{crit}$ for $\vartheta=0$ and $\widehat{\rho}_{crit}$ for $\vartheta=\pi$) in the effective potential (\ref{effkk}) when $\ell=0$ and $r_s\ll r_a=r_\Lambda$.  Here, $\rho=r/r_s$ and $x=r_s/r_a$ while $\rho_H$ and $\rho_h$ are computed from (\ref{radice1}) and (\ref{radice2}).}
\begin{center}
\begin{tabular}{ | l | l | l | l|l|l|l|}
\hline
$x$       & $\rho_H$  & $\rho_h$      &  $\rho_{crit}$ & $\widehat{\rho}_{crit}$ \\ \hline
$10^{-3}$ & 1         & 706.857       &  22.606        & 499.875\\ \hline
$10^{-4}$ & 1         & 7070.817      &  70.959        & 4999.875\\ \hline
$10^{-5}$ & 1         & 70710.428     &  223.856       & 49999.875\\ \hline
\end{tabular}
\label{tablerhocrit}
\end{center}
\end{table}
In order to find an analytic expression for the maximum by applying the perturbative theory of algebraic equations, it is convenient to use a different rescaling, namely $\widetilde{\rho}=\rho/r_\Lambda$. Then, $\rho=\widetilde{\rho}/x$ and the polynomial equation (\ref{cubo1}) becomes 
\begin{equation}\label{45}
(x+4)\widetilde{\rho}^3+(2-x)\widetilde{\rho}^2-3x\widetilde{\rho}-x=0.
\end{equation}
Since $x$ is a small parameter and the associated unperturbed polynomial has roots at $\widetilde{\rho}=-1/2,0,0$ a straightforward
application of perturbation methods for algebraic equations \cite{Murd} shows that (\ref{45}) has roots at
\begin{equation}
\widetilde{\rho}_1=\sqrt{\frac{x}{2}}+\frac{x}{4}-\frac{3\sqrt{2}}{32}x^{3/2}+\mathcal{O}(x^2),\quad
\widetilde{\rho}_2=-\frac{1}{2}-\frac{x}{8}+\mathcal{O}(x^2),\quad
\widetilde{\rho}_3=-\sqrt{\frac{x}{2}}+\frac{x}{4}+\frac{3\sqrt{2}}{32}x^{3/2}+\mathcal{O}(x^2).
\end{equation}
Moreover, Descartes' rule of signs implies that that there is only one positive root because the polynomial (\ref{45}) exhibits only one sign change due to the fact that $x\ll 1$ ensures that the term $2-x$ is  positive. Hence, we can conclude that $\widetilde{\rho}_{2,3}$ are negative and the only positive critical point is represented by the root $\widetilde{\rho}_1$. From case 3. in Section II the cosmological
horizon is located at $\widetilde{\rho}_h=(1/\sqrt{2})-(x/4)+\mathcal{O}(x^2)$. On the other hand, we find at quadratic order in $x$ that $\widetilde{\rho}_1<\widetilde{\rho}_h$ if $x\in(0,0.5046)$ while $\widetilde{\rho}_1>\widetilde{\rho}_H$ for $x\in(0,0.6774)$. Since $x\ll 1$, we conclude that the critical point is given by
\begin{equation}
r_{crit}=\sqrt{\frac{r_s r_\Lambda}{2}}+\frac{r_s}{4}-\frac{3\sqrt{2}}{32}r_s\sqrt{\frac{r_s}{r_\Lambda}}+\mathcal{O}\left(\frac{r_s^2}{r_\Lambda}\right)
\end{equation}
and by the analysis we performed previously, it must be a maximum for the effective potential.

\subsubsection{Time-like radial geodesics along $\vartheta=\pi$}
In this scenario, the corresponding cubic equation is
\begin{equation}\label{cubo2}
x^2(4-x)\rho^3-x(x+2)\rho^2+3x\rho-1=0.
\end{equation}
If we apply Descartes' rule of signs, we conclude that there are always 2 complex conjugate roots and one positive real root, here denoted by $\widehat{\rho}_{crit}$ because the polynomial (\ref{cubo2}) exhibits three sign changes due to the fact that $x\ll 1$ makes the term $4-x$ positive. Moreover,
\begin{equation}
\left.\frac{d^2\mathcal{V}_{eff}(\rho,\pi)}{d\rho^2}\right|_{\rho=\widehat{\rho}_{crit}}=-\frac{1}{\widehat{\rho}^3_{crit}}+x^2+\mathcal{O}\left(x^3\right)
\end{equation}
from which we conclude that $\widehat{\rho}_{crit}$ is not an equilibrium point for time-like particles moving along the south pole. Hence, a small perturbation will cause the particle to be either swallowed by the event horizon or to approach the cosmological horizon. For typical values of $\widehat{\rho}_{crit}$ we refer to Table~\ref{tablerhocrit}. Also in this case it possible to obtain an analytical expression for the maximum of the effective potential. Proceeding as before, we can rewrite (\ref{cubo2}) as
\begin{equation}\label{cubo22}
(x-4)\widetilde{\rho}^3+(2-x)\widetilde{\rho}^2-3x\widetilde{\rho}-x=0
\end{equation}
which has been obtained from (\ref{cubo2}) by setting $\widetilde{\rho}=\rho/r_\Lambda$ so that $\rho=\widetilde{\rho}/x$. The unperturbed polynomial has roots at $1/2$, $0$, $0$ and if we apply perturbative methods, it can be easily verified that
there are two complex conjugate roots and one real root given by $\widetilde{\rho}_{crit}=(1/2)-(x/8)+\mathcal{O}(x^2)$. In particular, we have
$\widetilde{\rho}_{crit}<\widetilde{\rho}_{h}$ if $x<1.6568$ and $\widetilde{\rho}_{crit}>\widetilde{\rho}_{H}$ for $x<0.4444$. 
Since $x\ll 1$, we conclude that $\widetilde{\rho}_{H}<\widetilde{\rho}_{crit}<\widetilde{\rho}_{h}$. Finally, we find that
\begin{equation}
r_{crit}=\frac{r_\Lambda}{2}-\frac{r_s}{8}+\mathcal{O}\left(\frac{r_s^2}{r_\Lambda}\right).
\end{equation}

\subsubsection{The case $\ell\neq 0$}
\begin{figure}[ht]
\includegraphics[width=0.5
\textwidth]{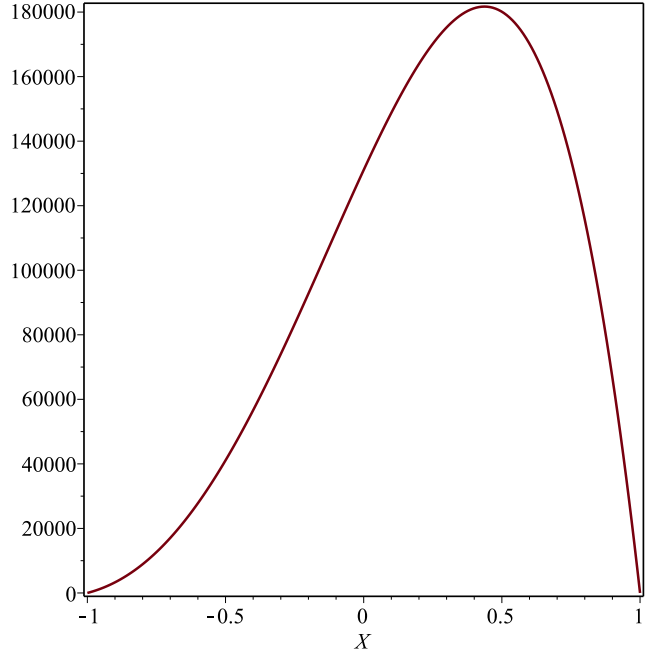}
\caption{Plot of the polynomial $P(X)$ defined by (\ref{pol}) with $X=\cos{\vartheta}$.}
\label{polinomio}
\end{figure}

For $\ell\neq 0$, the rescaled effective potential in the case $r_s\ll r_a=r_\Lambda$ reads
\begin{equation}\label{effuu}
\mathcal{U}_{eff}(\rho,\vartheta)=\frac{\left(1-\frac{1}{\rho}\right)(1-x^2\rho^2)-x^2\rho^2}{2\left(1+x\rho\cos{\vartheta}\right)^2}\left[1+\frac{L^2(1+x^2)\left(1+x\rho\cos{\vartheta}\right)^2}{\rho^2\left(1+x\cos{\vartheta}\right)\sin^2{\vartheta}}\right]
\end{equation}
with $L=\ell/r_\Lambda$, $x=r_s/r_\Lambda$ and $\rho=r/r_s$. Concerning the behaviour of the effective potential at the cosmological horizon, we observe that at the quadratic order in the small parameter $x$ 
\begin{equation}
\mathcal{U}_{eff}(\rho_h,\vartheta)=-\frac{114688\sin^2{\vartheta}}{P(X)}x^2+\mathcal{O}(x^3)
\end{equation}
with $X=\cos{\vartheta}$ and 
\begin{equation}\label{pol}
P(X)=-65536X^4-131072\sqrt{2}X^3-65536X^2+131072\sqrt{2}X+131072.
\end{equation}
At this point a comment is in order. Since the polynomial in (\ref{pol}) has roots at $\pm 1,\pm\sqrt{2}$, the potential will diverge at the cosmological horizon along the rays $\vartheta=0$ and $\vartheta=\pi$, i.e. along the direction of the cosmic string. On the other hand, for $\vartheta\in(0,\pi)$, the polynomial function $P(X)$ is always positive as it can be seen from Fig.~\ref{polinomio} and therefore, we conclude that for each fixed value of $\vartheta\in(0,\pi)$ the effective potential takes on a negative value at the cosmological horizon.
In the following, we perform a numerical analysis of the critical points of $\mathcal{U}_{eff}$. Imposing $\partial_\rho\mathcal{U}_{eff}=0$ and $\partial_\vartheta\mathcal{U}_{eff}=0$ leads to the following coupled system of algebraic equations
\begin{eqnarray}
&&\sum_{n=0}^5A_n(\vartheta){\rho}^{n}=0,\label{a1}\\
&&\sin{\vartheta}(2x^2\rho^3-x^2\rho^2-\rho+1)\sum_{n=0}^3 B_n(\vartheta)\rho^n=0\label{a2}
\end{eqnarray}
with
\begin{eqnarray}
A_5(\vartheta)&=&-x^4\cos^4{\vartheta}+x^3(L^2 x^4+L^2 x^2-5)\cos^3{\vartheta}+x^2(x^2-4)\cos^2{\vartheta}+5x^3\cos{\vartheta}+4x^2,\\
A_4(\vartheta)&=&-2x^2\cos^4{\vartheta}+x[2L^2x^2(x^2+1)-2]\cos^3{\vartheta}+3x^2[L^2x^2(x^2+1)+1]\cos^2{\vartheta}+x(2-x^2)\cos{\vartheta}-x^2,\\
A_3(\vartheta)&=&3x^2\cos^4{\vartheta}-3x[L^2 x^2(x^2+1)-1]\cos^3{\vartheta}+3x^2[2L^2(x^2+1)-1]\cos^2{\vartheta}+3x[L^2x^2(x^2+1)-1]\cos{\vartheta},\\
A_2(\vartheta)&=&x\cos^3{\vartheta}+-[9L^2 x^2(x^2+1)-1]\cos^2{\vartheta}+x[6L^2(x^2+1)-1]\cos{\vartheta}+L^2 x^2(x^2+1)-1,\\
A_1(\vartheta)&=&-9L^2 x(x^2+1)\cos{\vartheta}+2L^2(x^2+1)-3L^2(x^2+1),\quad
A_0(\vartheta)=-3xL^2(x+1)^2,\\
B_3(\vartheta)&=&-2x^3\cos^6{\vartheta}+x^2[3L^2 x^2(x^2+1)-4]\cos^5{\vartheta}+2x[L^2 x^2(x^2+1)+2x^2-1]\cos^4{\vartheta}\\
&&-x^2[L^2x^2(x^2+1)-8]\cos^3{\vartheta}-2x(x^2-2)\cos^2{\vartheta}-4x^2\cos{\vartheta}-2x,\quad
B_2(\vartheta)=x\cos{\vartheta}B_1(\vartheta),\\
B_1(\vartheta)&=&9L^2 x^2(x^2+1)\cos^3{\vartheta}+6L^2 x(x^2+1)\cos^2{\vartheta}-3L^2 x^2(x^2+1)\cos{\vartheta},\\
B_0(\vartheta)&=&3L^2 x(x^2+1)\cos^2{\vartheta}+2L^2(x^2+1)\cos{\vartheta}-L^2 x(x^2+1).
\end{eqnarray}
First of all, we observe that even though $\vartheta=0,\pi$ are roots for the equation (\ref{a2}), they must be disregarded because the effective potential is singular there. Concerning the roots of the polynomial $\mathfrak{p}(x)=2x^2\rho^3-x^2\rho^2-\rho+1$, Descartes' rule of signs  signalizes the presence of two or zero positive roots. However, these roots are not relevant to the present analysis because they coincide with the event and cosmological horizons. This can be easily seen by rewriting $\mathfrak{p}(x)=0$ as the cubic equation (\ref{pertur}) and taking into account that $\epsilon=x$. These observations tell us that it suffices to consider the following system
\begin{equation}
\sum_{n=0}^5A_n(\vartheta){\rho}^{n}=0,\quad
\sum_{n=0}^3 B_n(\vartheta)\rho^n=0.
\end{equation}
In the Table~\ref{tablerhocritL}, we classified the critical points of the effective potential (\ref{effuu}) for $x$ in the range $10^{-5}\div 10^{-3}$ and $L$ between $10^{-2}$ and $1.733$. We observe that for small values of $L$ the potential admits only saddle points. However, as $L$ increases, a local minimum develops even if the values of the parameter $x$ decreases. In Table~\ref{tablerhocritLL}, we focus on the dynamics of the local minimum when $L$ increases while $x$ remains fixed. More precisely, a local minimum exists only if $L$ varies between some $L_{min}$ and $L_{max}$. In particular for $x=10^{-3}$, we find $L_{min}\approx 1.7317$ and $L_{max}\approx 2.4380$.
\begin{table}[ht]
\caption{Typical values for the saddle points and local minima of the effective potential (\ref{effkk}) when $\ell\neq 0$, $10^{-5}\leq x\leq 10^{-3}$ and $10^{-2}\leq L\leq 1.733$. Here, "sp" and "lm" stand for saddle point and local minimum, respectively.}
\begin{center}
\begin{tabular}{ | l | l | l | l|l|l|l|}
\hline
$x$       & $L$  & $\rho_{crit}$     &  $\vartheta_{crit}$ (rad) & Type \\ \hline
$10^{-3}$ & $10^{-2}$      & 22.591       &  0.055                    & sp\\ \hline
$10^{-4}$ & "         & 70.930       &  0.041                    & sp\\ \hline
$10^{-5}$ & "         & 223.803      &  0.031                    & sp\\ \hline
$10^{-3}$ & $10^{-1}$       & 22.449       &  0.175                    & sp\\ \hline
$10^{-4}$ & "         & 70.666       &  0.131                    & sp\\ \hline
$10^{-5}$ & "         & 223.329      &  0.098                    & sp\\ \hline
$10^{-3}$ & 1         & 20.764       &  0.593                    & sp\\ \hline
$10^{-4}$ & "         & 67.776       &  0.428                    & sp\\ \hline
$10^{-5}$ & "         & 218.340      &  0.315                    & sp\\ \hline
$10^{-3}$ & 1.731     & 18.741       &  0.845                    & sp\\ \hline
$10^{-4}$ & "         & 64.980       &  0.584                    & sp\\ \hline
$10^{-5}$ & "         & 213.917      &  0.422                    & sp\\ \hline
$10^{-3}$ & 1.7317    & 2.998        &  1.561                    & sp\\ \hline
"         & "         & 3.007        &  1.561                    & lm\\ \hline
"         & "         & 18.738       &  0.845                    & sp\\ \hline
$10^{-4}$ & "         & 64.977       &  0.584                    & sp\\ \hline
$10^{-5}$ & "         & 213.912      &  0.422                    & sp\\ \hline
$10^{-3}$ & 1.733     & 2.892        &  1.562                    & sp\\ \hline
"         & "         & 3.122        &  1.560                    & lm\\ \hline
"         & "         & 18.734       &  0.846                    & sp\\ \hline
$10^{-4}$ & "         & 2.90377      &  1.56999                  & sp\\ \hline
"         & "         & 3.10287      &  1.56975                  & lm\\ \hline
"         & "         & 64.972       &  0.585                    & sp\\ \hline
$10^{-5}$ & "         & 2.90390      &  1.57071                  & sp\\ \hline
"         & "         & 3.10267      &  1.57069                  & lm\\ \hline
"         & "         & 213.904      &  0.422                    & sp\\ \hline
\end{tabular}
\label{tablerhocritL}
\end{center}
\end{table}
\begin{table}[ht]
\caption{Typical values for the saddle points and local minima of the effective potential (\ref{effkk}) when $\ell\neq 0$, $x=10^{-3}$ and $1.8\leq L\leq 100$. As in the previous table, sp=saddle point and lm=local minimum.}
\begin{center}
\begin{tabular}{ | l | l | l | l|l|l|l|}
\hline
$L$      & $\rho_{crit}$     &  $\vartheta_{crit}$ (rad) & Type \\ \hline
1.8      & 2.358             &  1.566                    & sp\\ \hline
"        & 4.136             &  1.548                    & lm\\ \hline
"        & 18.488            &  0.871                    & sp\\ \hline
2.2      & 1.856             &  0.157                    & sp\\ \hline
"        & 8.115             &  1.462                    & lm\\ \hline
"        & 16.499            &  1.042                    & sp\\ \hline
2.4      & 1.773             &  1.569                    & sp\\ \hline
"        & 11.059            &  1.356                    & lm\\ \hline
"        & 14.406            &  1.185                    & sp\\ \hline
2.438    & 1.761             &  1.569                    & sp\\ \hline
"        & 12.492            &  1.290                    & lm\\ \hline
"        & 13.131            &  1.257                    & sp\\ \hline
2.440    & 1.760             &  1.569                    & sp\\ \hline
10       & 1.511             &  1.570                    & lm\\ \hline
$10^2$      & 1.500             &  1.570                    & sp\\ \hline
\end{tabular}
\label{tablerhocritLL}
\end{center}
\end{table}
\begin{figure}[ht]
\includegraphics[width=0.5
\textwidth]{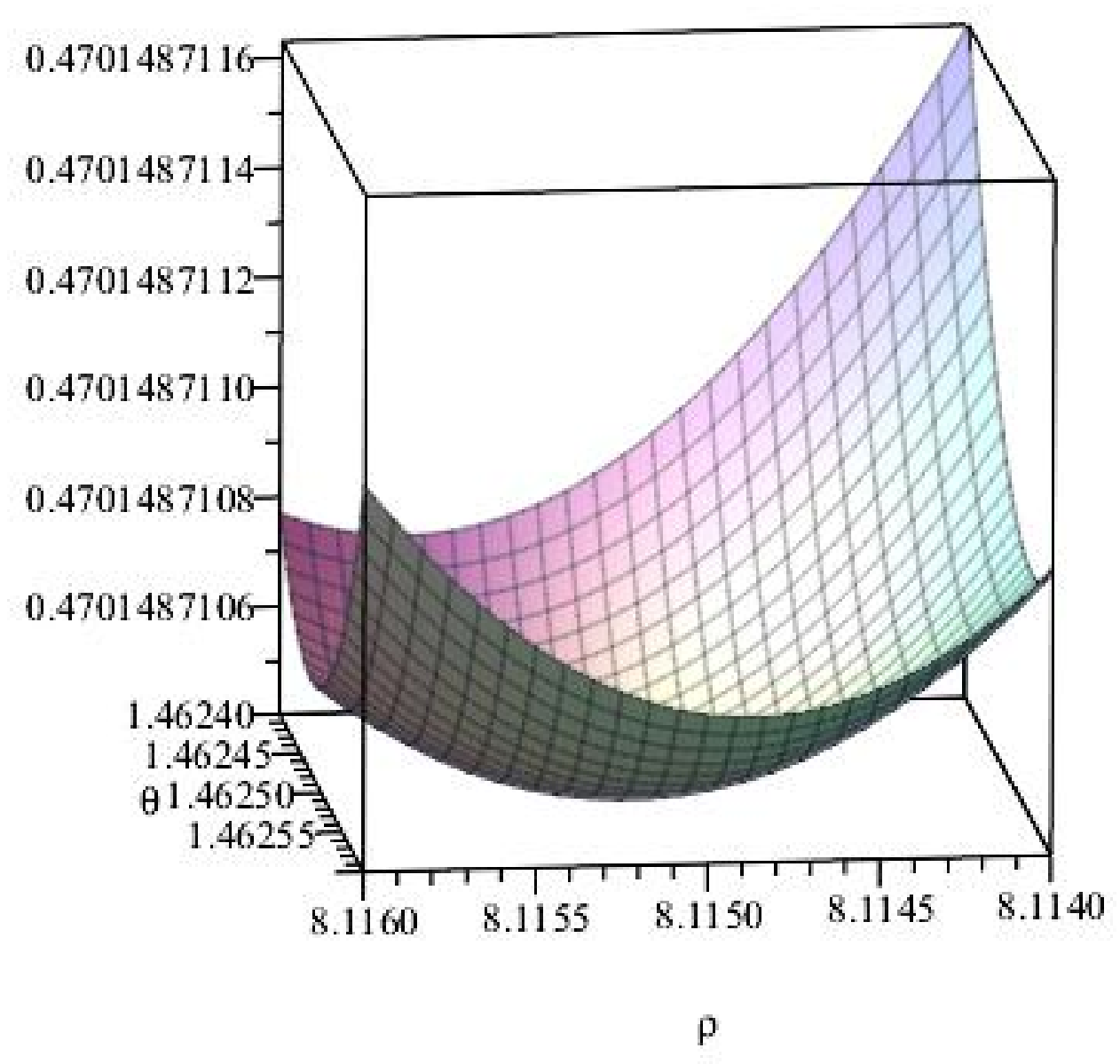}
\caption{Local minimum for the potential (\ref{effuu}) for $L=2.2$ and $x=10^{-3}$. The minimum is located at $(\rho_m,\vartheta_m)=(8.11523,1.46250)$ where $\mathcal{U}_{eff}(\rho_m,\vartheta_m)\approx 0.4701487104$.}
\label{min}
\end{figure}
One can see from Table~\ref{tablerhocritLL} the prevalence of saddle points. It is tempting to call the effective potential of the C-metric with positive cosmological constant the potential of saddle points. The latter is positive and becomes negative at large $\rho$ (see equation  (\ref{large})). This achieved by a saddle point as we demonstrate in Fig.~\ref{eccebombo}.
\begin{figure}[ht]
\includegraphics[width=0.5
\textwidth]{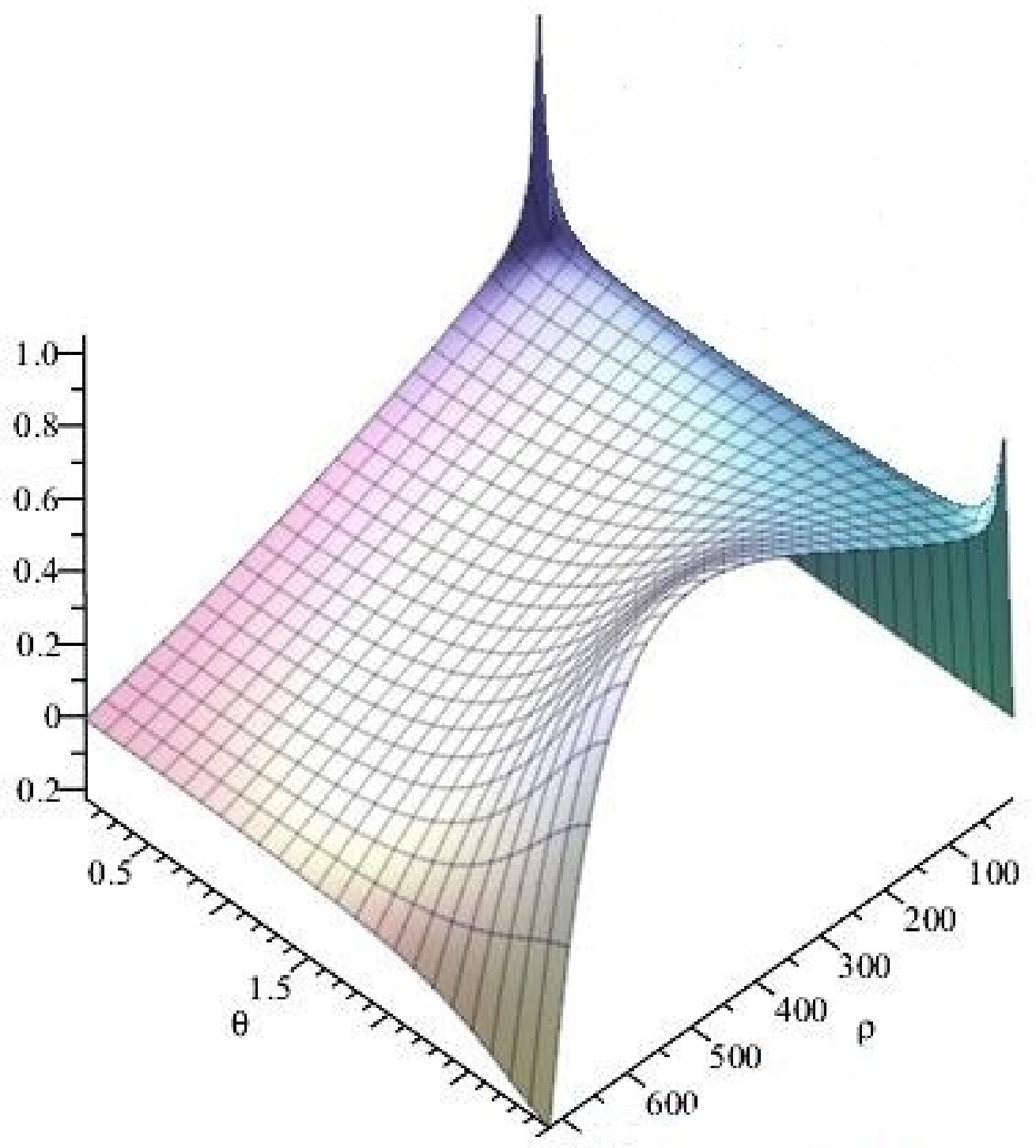}
\caption{Typical saddle point configuration for the potential (\ref{effuu}) for $L=2.2$ and $x=10^{-3}$.}
\label{eccebombo}
\end{figure}

\section{Jacobi stability analysis of the null circular Orbits}
Since we do not know a priori which effect $\Lambda$ has on the stability of the null circular orbits found in the previous section, we need to study once again the salient features of the associated Jacobi stability problem. To this purpose, we need first to verify that the critical point of the effective potential is also a critical point for the geodesic equations (\ref{g7}) and (\ref{g8}). In that regard, it is convenient to rewrite (\ref{g7}) and (\ref{g8}) with the help of the constraint equation (\ref{cos}) and the definition of the effective potential (\ref{Uefflight}) as follows
\begin{eqnarray}
&&\frac{d^2 r}{d\lambda^2}+[\partial_r\ln{\sqrt{A_\Lambda C}}]\left(\frac{dr}{d\lambda}\right)^{2}+(\partial_\vartheta\ln{A_\Lambda}) \frac{dr}{d\lambda}\frac{d\vartheta}{d\lambda}+\frac{E}{F^2}\partial_r\ln{\frac{B_\Lambda}{C}}=0,\label{din1}\\
&&\frac{d^2 \vartheta}{d\lambda^2}+[\partial_\vartheta\ln{\sqrt{A_\Lambda C}}]\left(\frac{d\vartheta}{d\lambda}\right)^{2}+(\partial_r\ln{C})\frac{dr}{d\lambda}\frac{d\vartheta}{d\lambda}+\frac{\ell^2}{2CD}\partial_\vartheta\ln{\frac{A_\Lambda}{D}}=0.\label{din2}
\end{eqnarray}
For a circular orbit with $r=r_c$ and $\vartheta=\vartheta_c$ all derivatives in the above equations vanish and we are left with the following system of equations
\begin{equation}
\left.\partial_r\left(\frac{B_\Lambda}{C}\right)\right|_{(r_c,\vartheta_c)}=0,\quad
\left.\partial_\vartheta\left(\frac{A_\Lambda}{D}\right)\right|_{(r_c,\vartheta_c)}=0,
\end{equation}
that can be simplified as follows
\begin{equation}\label{radius}
\left.\frac{d}{dr}\left(\frac{f_\Lambda(r)}{r^2}\right)\right|_{r=r_c}=0,\quad
\left.\frac{d}{d\vartheta}\left(\frac{1}{g\sin^2{\vartheta}}\right)\right|_{\vartheta=\vartheta_c}=0.
\end{equation}
Since 
\begin{equation}
\partial_r \mathfrak{V}=\frac{\ell^2}{2\kappa^2 g\sin^2{\vartheta}}\frac{d}{dr}\left(\frac{f_\Lambda(r)}{r^2}\right),\quad
\partial_\vartheta \mathfrak{V}=\frac{\ell^2 f_\Lambda}{2\kappa^2 r^2}\frac{d}{d\vartheta}\left(\frac{1}{g\sin^2{\vartheta}}\right)
\end{equation}
and the derivatives above vanish when evaluated at $r=r_c$ and $\vartheta=\vartheta_c$, we conclude that the equations in (\ref{radius}) are trivially satisfied. In order to study the Jacobi (in)stability of the null circular orbits, we will proceed as in \cite{Maha}, that is, we first recognize that the equations (\ref{din1}) and (\ref{din2}) are a special case of the dynamical system
\begin{equation}\label{sist}
\frac{d^2 x^i}{d\lambda^2}+g^i(x^1,x^2,y^1,y^2)=0
\end{equation}
where
\begin{eqnarray}
g^1(x^1,x^2,y^1,y^2)&=&[\partial_1\ln{\sqrt{A_\Lambda C}}](y^1)\strut^{2}+(\partial_2\ln{A_\Lambda}) y^1y^2+\frac{E}{F^2}\partial_1\ln{\frac{B_\Lambda}{C}},\label{gg1}\\
g^2(x^1,x^2,y^1,y^2)&=&[\partial_2\ln{\sqrt{A_\Lambda C}}](y^2)\strut^{2}+(\partial_1\ln{C}) y^1y^2+\frac{\ell^2}{2CD}\partial_2\ln{\frac{A_\Lambda}{D}}\label{gg2}
\end{eqnarray}
with $x^1:=r$, $x^2:=\vartheta$, and $y^i=dx^{i}/d\lambda$ for $i=1,2$ and then, we apply the Kosambi-Cartan-Chern (KCC) theory which has been widely used in the last decade as a powerful toll to probe the stability of several dynamical systems appearing in gravitation and cosmology  \cite{21,22,23,28,H,Baha,Bom,Bom1,Harko1,Danila,Lake}. Let us assume that $g^1$ and $g^2$ are smooth functions in a neighbourhood of the initial condition $(x^1_0,x^2_0,y^1_0,y^2_0,\lambda_c)=(r_c,\vartheta_c,0,0,\lambda_c)\in\mathbb{R}^5$. The main result we will use is the following theorem: {\it{an integral curve $\gamma$ of (\ref{sist}) is Jacobi stable if and only if the real parts of the eigenvalues of the second KCC invariant $P^i_j$ are strictly negative everywhere along $\gamma$, and Jacobi unstable otherwise}}. We recall that 
\begin{equation}\label{KKC2}
P^i_j=-\frac{\partial g^i}{\partial x^j}-g^rG^i{}_{rj}+y^r\frac{\partial N^i_j}{\partial x^r}+N^i_r N^r_j+\frac{\partial N^i_j}{\partial\lambda},\quad G^i{}_{rj}=\frac{\partial N^i_r}{\partial y^j},\quad N^i_j=\frac{1}{2}\frac{\partial g^i}{\partial y^j},
\end{equation}
where $G^i{}_{rj}$ is called the Berwald connection \cite{Anto,Miron}. Observe that the term $\partial N^i_j/\partial\lambda$ in (\ref{KKC2}) does not contribute because the system (\ref{sist}) is autonomous in the variable $\lambda$. For a proof of the above result we refer to \cite{1,6,Bom}. In preparation to the application of this theorem, we introduce the matrix associated to the second KCC invariant, namely
\begin{equation}\label{Matrize}
\widetilde{P}:=\left(
\begin{array}{cc}
\widetilde{P}^1_1 & \widetilde{P}^1_2\\
\widetilde{P}^2_1 & \widetilde{P}^2_2
\end{array}
\right),
\end{equation}
where a tilde means evaluation at $x^1=r_c$ and $x^2=\vartheta_c$. The associated characteristic equation for the eigenvalues is
\begin{equation}\label{cee}
\mbox{det}\left(
\begin{array}{cc}
\widetilde{P}^1_1-\lambda & \widetilde{P}^1_2\\
\widetilde{P}^2_1 & \widetilde{P}^2_2-\lambda
\end{array}
\right)=0.
\end{equation}
First of all, we observe that $y^{i}$ with $i=1,2$ vanishes along the null circular orbit. This implies that the third term on the r.h.s. of the first equation in (\ref{KKC2}) does not give any contribution. By the same token, $N^i_j$ defined in (\ref{KKC2}) depends quadratically in $y^1$ and $y^2$ and hence, its first order partial derivatives with respect to $y^i$ are linear combinations in $y^i$ vanishing once evaluated at $x^1=r_c$ and $x^2=\vartheta_c$. Hence, we have 
\begin{equation}
\widetilde{P}^i_j=-\left(\frac{\partial g^i}{\partial x^j}+\frac{1}{2}g^r\frac{\partial^2 g^i}{\partial y^j\partial y^r}\right)_{x^1=r_c,~x^2=\vartheta_c}.
\end{equation}
After a lengthy but straightforward computation we find that
\begin{eqnarray}
\widetilde{P}^1_2&=&-E\left[\frac{\partial}{\partial\vartheta}\left(\frac{C}{F^2 B_\Lambda}\frac{\partial}{\partial r}\left(\frac{B_\Lambda}{C}\right)\right)+\frac{C}{2F^4}\frac{\partial}{\partial r}\left(\frac{B_\Lambda}{C}\right)\frac{\partial A_\Lambda}{\partial\vartheta}\right]_{r=r_c,~\vartheta=\vartheta_c},\nonumber\\
&=&\frac{3E r^2_c}{2f_\Lambda(r_c)F^3(r_c,\vartheta_c)}\left[\frac{d}{dr}\left(\frac{f_\Lambda}{r^2}\right)\frac{\partial F}{\partial\vartheta}\right]_{r=r_c,~\vartheta=\vartheta_c}.
\end{eqnarray}
By means of the first equation in (\ref{radius}) we immediately conclude that $\widetilde{P}^1_2=0$. This implies that the eigenvalues of the matrix (\ref{Matrize}) are given by 
\begin{equation}
\lambda_1=\widetilde{P}^1_1,\quad\lambda_2=\widetilde{P}^2_2.
\end{equation}
Let us analyze the sign of $\lambda_1$. We observe that by means of  (\ref{radius})
\begin{eqnarray}
\lambda_1&=&-\left\{E\left[\frac{\partial}{\partial r}\left(\frac{C}{F^2 B_\Lambda}\frac{\partial}{\partial r}\left(\frac{B_\Lambda}{C}\right)\right)+\frac{1}{2F^4}\frac{\partial}{\partial r}\left(\frac{B_\Lambda}{C}\right)\frac{\partial A_\Lambda C}{\partial r}\right]+\frac{\ell^2}{4A_\Lambda^2 C}\frac{\partial}{\partial\vartheta}\left(\frac{A_\Lambda}{D}\right)\frac{\partial A_\Lambda}{\partial\vartheta}\right\}_{r=r_c,~\vartheta=\vartheta_c},\nonumber\\
&=&-\frac{Er_c^2}{f_\Lambda(r_c)F^2(r_c,\vartheta_c)}\left.\frac{d^2}{dr^2}\left(\frac{f_\Lambda}{r^2}\right)\right|_{r=r_c},\nonumber\\
&=&\frac{Er_c^3}{f_\Lambda(r_c)F^2(r_c,\vartheta_c)}\frac{1+12\alpha^2 M^2+\sqrt{1+12\alpha^2 M^2}}{1+12\alpha^2 M^2}.
\end{eqnarray}
Since the energy $E$ is positive, $r_H<r_c<r_h$ and $f_\Lambda$ is positive on the interval $(r_H,r_h)$, we conclude that $f_\Lambda(r_c)>0$ and the eigenvalue $\lambda_1$ is always strictly positive. This implies that a photon circular orbit with radius $r_c$ on the cone $\vartheta=\vartheta_c$ are Jacobi unstable. 

\section{Gravitational lensing}
The cosmological constant seems to be an obstacle in calculating the deflection angle of light in a curved spacetime as it can be evinced from \cite{Arakida} where a comparison of the different results in the Schwarschild-de Sitter spacetime has been provided. It is therefore of some interest to perform the calculation of light deflection in the C-metric with $\Lambda$. We recall that distance measures, image distortion and image brightness of an astrophysical object  hidden by a gravitational lens require the analysis of the equation of geodesic deviation and in particular the derivation of the so-called Sachs optical scalars allowing to study the null geodesic congruences \cite{Sachs,Wald,Seitz}. Without much further ado we observe that a construction of a symmetric null tetrad in the spirit of \cite{Carter} (see equation (5.119) therein) can be performed as in \cite{Maha} by replacing there the function $f$ by $f_\Lambda$ given in (\ref{Ffg}). However, this procedure leads to a non-vanishing spin coefficient $\epsilon$, thus signalizing that the null geodesics are not affinely parameterized. The solution to this problem consists in realizing that the spin coefficient $\kappa$ is zero also in the case of the metric (\ref{metricFinal}) and therefore, the construction of an affine parameterization can be achieved in terms of a rotation of class III (see $\S$7(g) eq. (347) in \cite{CH}) which preserves the direction of the tetrad basis vector $\bm{\ell}$ while keeping $\kappa=0$. As a consequence of this approach, the spin coefficient $\sigma$ will provide access to the Sachs optical scalar describing the shear effect on the light beam due to the gravitational field. To this purpose, we consider the normalized null tetrad $(\bm{\ell},\mathbf{n},\mathbf{m},\overline{\mathbf{m}})$
\begin{equation}\label{new tetrad}
\ell_{i}=\left(\frac{1}{\sqrt{2}},\frac{1}{\sqrt{2}f_\Lambda},0,0\right),\quad
n_{i}=\left(\frac{f_\Lambda F}{\sqrt{2}},-\frac{F}{\sqrt{2}},0,0\right),\quad 
m_{i}=\left(0,0,r\sqrt{\frac{F}{2g}},i\kappa r\sin{\vartheta}\sqrt{\frac{Fg}{2}}\right)
\end{equation}
and we recall that in the Newmann-Penrose formalism the ten independent components of the Weyl tensor are replaced by five scalar fields $\Psi_0,\cdots,\Psi_4$ while the ten components of the Ricci tensor are expressed in terms of the scalar fields $\Phi_{ab}$ with $a,b=0,1,2$ and the Ricci scalar $R$ is written by means of the scalar field $\widehat{\Lambda}=R/24$. The spin coefficients for our problem are computed to be 
\begin{eqnarray}
\kappa&=&\sigma=\lambda=\nu=\epsilon=0,\quad
\rho=\frac{1}{r\sqrt{2F}},\quad\mu=\frac{f_\Lambda}{r}\sqrt{\frac{F}{2}},\quad
\tau=-\pi=\frac{1}{2rF}\sqrt{\frac{g}{2F}}\frac{\partial F}{\partial\vartheta},\\
\gamma&=&-\frac{1}{2\sqrt{2}F}\frac{\partial(f_\Lambda F)}{\partial r},\quad
\beta=-\frac{1}{2\sqrt{2}rF\sin{\vartheta}}\frac{\partial \sqrt{Fg}\sin{\vartheta}}{\partial\vartheta}+\frac{\tau}{2},\quad
\alpha=\frac{1}{2\sqrt{2}rF\sin{\vartheta}}\frac{\partial \sqrt{Fg}\sin{\vartheta}}{\partial\vartheta}+\frac{\tau}{2}.\label{aa}
\end{eqnarray}
and the only non-vanishing scalar fields $\Psi_i$, $\Phi_{ab}$, and $\widehat{\Lambda}$ for a two black hole metric with positive cosmological constant are
\begin{eqnarray}
\Psi_2&=&\frac{1}{3}\left[\ell^r\partial_r\mu+n^r\partial_r\gamma-m^\vartheta\partial_\vartheta(\pi+\alpha)+\overline{m^\vartheta}\partial_\vartheta\beta+(\alpha-\beta)(\alpha-\beta+\pi)\right],\\
\Phi_{11}&=&\frac{1}{2}\left[n^r\partial_r\gamma+m^\vartheta\partial_\vartheta\alpha-\overline{m^\vartheta}\partial_\vartheta\beta+\tau^2-\mu\rho-(\alpha-\beta)^2\right],\\
\widehat{\Lambda}&=&\Psi_2-\Phi_{11}+m^\vartheta\partial_\vartheta\alpha-\overline{m^\vartheta}\partial_\vartheta\beta-\mu\rho-(\alpha-\beta)^2.
\end{eqnarray}
At this point a remark is in order. First of all, the cosmological constant enters only in the spin coefficients $\mu$ and $\gamma$ while the other spin coefficients are the same as those obtained for the $C$-metric in \cite{Maha}. Moreover, the fact that $\rho$ is real has a twofold implication: the congruence of null geodesics is hypersurface orthogonal and accordingly, the optical scalar $\omega=\Im{\rho}$ must vanish. In other words, a light beam propagating in the metric described by (\ref{metricFinal}) does not get twisted or rotated. Furthermore, if we consider equations ($310$a) and ($310$b) (see $\S$8(d) p. 46 in \cite{CH}) describing how the spin coefficients $\rho$ and $\sigma$ vary along the geodesics
\begin{eqnarray}
D\rho&=&\rho^2+|\sigma|^2+\Phi_{00},\quad D=\ell^a\partial_a\label{eins}\\
D\sigma&=&2\sigma\rho+\Psi_0,\label{zwei}
\end{eqnarray}
we immediately observe that the second equation is of no practical use because it is always trivially satisfied ($\sigma=0=\Psi_0$). In addition, the vanishing of the spin coefficient $\sigma$ is signalizing that a light beam does not experience any shear effect, i.e. if the light beam has initially a circular cross section such a cross section does not change its shape after the interaction with the black hole took place. Finally, the optical scalar $\theta$ which measures the contraction/expansion of a light beam travelling through the given gravitational field, is expressed in terms of the spin coefficient $\rho$ as
\begin{equation}
\theta=-\Re{\rho}=-\frac{1}{r\sqrt{2F}}.
\end{equation}
The fact that $\theta$ is negative implies that the light beam undergoes a compression process in the presence of a two black hole metric with positive cosmological constant. However, as $\theta$ does not depend on $\Lambda$ and it coincides with the corresponding optical scalar computed for the $C$-metric in \cite{Maha}, it is impossible to distinguish a $C$-black hole from the one described by (\ref{metricFinal}) if we limit us to 
probe only into effects in the optical scalar $\theta$. This observation suggests that we need to study the weak and strong gravitational lensing in order to detect some distinguishing features among the aforementioned black hole solutions. We start by observing that in our situation the weak lensing problem can be tackled by a method similar to that adopted in \cite{Maha} due to the fact that the saddle point $(r_c,\vartheta_c)$ of the effective potential (\ref{Uefflight}) coincides with the critical point of the dynamical system (\ref{g7})-(\ref{g8}). As in \cite{Maha}, we will assume that the light ray and the observer are positioned on the cone $\vartheta=\vartheta_c$. Then, the angular motion is controlled by the equations
\begin{equation}\label{bubu}
\frac{d\phi}{d\lambda}=\frac{\ell}{D(r,\vartheta_c)},\quad\vartheta=\vartheta_c,
\end{equation}
the time-like variable $t$ is linked to the parameterization $\lambda$ according to
\begin{equation}
\frac{dt}{d\lambda}=\frac{\mathcal{E}}{B_\Lambda(r,\vartheta_c)},
\end{equation}
while the radial motion is described by the following equation obtained by combining (\ref{cos}) with (\ref{Uefflight}), namely
\begin{equation}\label{25.19}
\left(\frac{dr}{d\lambda}\right)^2=\frac{1}{A_\Lambda(r,\vartheta_c)}\left[\frac{\mathcal{E}^2}{B_\Lambda(r,\vartheta_c)}-\frac{\ell^2}{D(r,\vartheta_c)}\right].
\end{equation}
In order to determine the trajectory $\phi=\phi(r)$ on the cone $\vartheta=\vartheta_c$, a trivial application of the Chain Rule to $d\phi/d\lambda$ combined with (\ref{25.19}) leads to 
\begin{equation}\label{25.20}
\frac{d\phi}{dr}=\frac{F(r,\vartheta_c)}{\sqrt{D(r,\vartheta_c)}}
\left[\frac{\mathcal{E}^2}{\ell^2}D(r,\vartheta_c)-B_\Lambda(r,\vartheta_c)\right]^{-1/2},
\end{equation}
where without loss of generality we picked the plus sign corresponding to a null ray approaching the black hole along an anticlockwise trajectory. Moreover, like in \cite{Weinberg}, the quantity $\mathcal{E}/\ell$ has the interpretation of an impact parameter $b$ defined as
\begin{equation}\label{impact}
\frac{1}{b^2}=\frac{\mathcal{E}^2}{\ell^2}=\frac{B_\Lambda(r_0,\vartheta_c)}{D(r_0,\vartheta_c)}=\frac{1}{\kappa^2 g(\vartheta_c)\sin^2{\vartheta_c}}\frac{f_\Lambda(r_0)}{r_0^2},
\end{equation}
where $r_0>r_H$ is the distance of closest approach and $r_H$ denotes the event horizon. In order to check the validity of (\ref{impact}), let us choose $r_0=r_c$ with $r_c$ denoting the radius of the circular orbits and recall that the critical impact parameter in the case of the Schwarzschild-de Sitter metric is given by \cite{usPRD}
\begin{equation}
\widetilde{b}_c=\frac{3\sqrt{3}M}{\sqrt{1-\frac{27}{4}y}},\quad y=\left(\frac{r_s}{r_\Lambda}\right)^2=\frac{4}{3}M^2\Lambda,
\end{equation}
where the gravitational lensing can only be studied for $0<y<4/27$ because as $y\to 4/27$ from the left the event and cosmological horizons of the Schwarzschild-de Sitter black hole would shrink and coalesce with the radius of the photon sphere at $r_\gamma=3M$. Then, the critical impact parameter $b_c$ can be obtained from (\ref{impact}) as
\begin{equation}\label{imp}
b_c=\kappa\sin{\vartheta_c}\sqrt{g(\vartheta_c)}\frac{r_c}{\sqrt{f_\Lambda(r_c)}}.
\end{equation}
Moreover, let us remind the reader that in the case of vanishing acceleration, i.e. $\alpha\to 0$, our metric goes over into the Schwarzschild-de Sitter metric. A Taylor  expansion of (\ref{imp}) around $x=0$ with $x=\alpha M$  leads to
\begin{equation}\label{exp}
b_c=\widetilde{b}_c-2\widetilde{b}_c x+\frac{\widetilde{b}_c}{6}(\widetilde{b}_c^2+27)x^2+\mathcal{O}\left(x^3\right).
\end{equation}
It is gratifying to observe that (\ref{exp}) correctly reproduces the Schwarzschild-de Sitter critical impact parameter in the limit $x\to 0$ while it also agrees for $y\to 0$ with the critical impact parameter for a $C$-black hole (see equation (93) in \cite{Maha}). Having determined the critical impact parameter for our problem allows to distinguish among the following scenarios
\begin{enumerate}
\item
if $b<b_c$, the photon is captured by the black hole;
\item
if $b>b_c$, deflection takes place and two further cases are possible, namely
\begin{enumerate}
\item
if $b\gg b_c$ or equivalently $r_0\gg r_c$, the trajectory is almost a straight line and we are in the regime of weak gravitational lensing.
\item
If $b\gtrsim b_c$ or equivalently $r_0\gtrsim r_c$, strong gravitational lensing occurs with the photon orbiting several times around the black hole before it flies off.
\end{enumerate}
\end{enumerate}
If we go back to (\ref{25.20}), we observe that the function $D(r,\vartheta_c)$ can never be negative while the same can not be said for the other square root. This means that some motion reality condition should be introduced. This can be easily done by rescaling the radial variable according to $\rho=r/r_s$ and setting $\widehat{x}=2\alpha M$ and $d=(r_s/r_\Lambda)^2$. Then, 
\begin{equation}
f_\Lambda(\rho)=1-\frac{1}{\rho}-\widehat{x}^2(\rho^2-\rho)-d\rho^2
\end{equation}
and by means of (\ref{impact}) and (\ref{coefficienti}) equation (\ref{25.20}) becomes
\begin{eqnarray}
\frac{d\phi}{d\rho}&=&\frac{\rho_0}{\kappa\rho\sqrt{g(\vartheta_c)}\sin{\vartheta_c}}
\frac{1}{\sqrt{\rho^2 f_\Lambda(\rho_0)-\rho_0^2 f_\Lambda(\rho)}},\label{opla1}\\
&=&\frac{\rho_0\sqrt{\rho_0}}{\kappa\sqrt{\rho g(\vartheta_c)}\sin{\vartheta_c}}\frac{1}{\sqrt{(\widehat{x}^2\rho_0^2+\rho_0-1)\rho^3-\rho_0^3(\widehat{x}^2\rho^2+\rho-1)}},\label{opla2}
\end{eqnarray}
where $\rho_0$ is the rescaled distance of closest approach. At this point, it is interesting to observe a couple of facts. First of all, there is no dependence on the cosmological constant in the expression above. This feature is already present in the Schwarzschild-de Sitter case \cite{usPRD} where $\Lambda$ can influence the trajectories of massive particles while it is absent in the coordinate orbital equation when photons are considered \cite{Islam}. However, the cosmological constant can appear in the formula for the deflection angle in the weak regime when the observer is close to the cosmological horizon. In addition to the previous remark, equation (\ref{opla2}) coincides with equation (98) in \cite{Maha}. This implies that the analysis of the turning points performed by \cite{Maha} for the $C$-metric will continue to hold also in the present case. For this reason, we will limit us to recall only those basic facts that are necessary in order to proceed further with the analysis of the weak/strong gravitational lensing. First of all, we remind the reader that under the assumption $\rho_s\ll\rho_\Lambda$ we have  $\rho_H>1$ and therefore, $\rho_0>\rho_H>1$. Moreover, the cubic equation in (\ref{opla2}) admits three real turning points, namely $\rho_0$ and $\rho_\pm$ where an analytic expression for $\rho_\pm$ is given by (100) in \cite{Maha}. When integrating (\ref{opla2}) is extremely important to know the spatial ordering of the points $\rho_H$, $\rho_0$, $\rho_\pm$, and $\rho_h$. For a proof of the results summarized here below we refer to Appendix C in \cite{Maha}. 
\begin{enumerate}
\item
Weak lensing: $\rho_0\gg\rho_c$. If $\rho_0>\rho_\gamma>\rho_c$ with $\rho_\gamma$ representing the radius of the Schwarzschild photon sphere, it follows that $\rho_+<\rho_c<\rho_0$ for any $\widehat{x}\in(0,1)$. This implies that $\rho_{-}<0<\rho_+<\rho_0$ and the cubic in (\ref{opla2}) is positive on the interval $(\rho_0,\rho_h)$.
\item
Strong lensing: $\rho_c\lesssim\rho_0<\rho_\gamma$ for $\widehat{x}\in(\sqrt{2(\rho_\gamma-\rho_0)}/\rho_0,1)$. If $\widehat{x}$ is in the aforementioned range, then  $\rho_+<\rho_0$ and $\rho_+<\rho_c$. This ensures that the cubic in (\ref{opla2}) is positive on the interval $(\rho_0,\rho_h)$.
\end{enumerate}
Let us focus on the weak gravitational lensing. By $\rho_b$ we denote the position of the observer which must be placed in the interval $(\rho_c,\rho_h)$. At this point, by means of the angular transformation $\phi=\varphi/k$ we can integrate (\ref{opla1}) and cast the integral into the form
\begin{equation}\label{soluza}
\varphi(\rho_0)=\frac{1}{\sqrt{g(\vartheta_c)}\sin{\vartheta_c}}\int_{\rho_0}^{\rho_b}\frac{d\rho}{\rho\sqrt{f_\Lambda(\rho)}}\left[\left(\frac{\rho}{\rho_0}\right)^2 \frac{f_\Lambda(\rho_0)}{f_\Lambda(\rho)}-1\right]^{-1/2}.
\end{equation}
We remind the reader that, unlike the Schwarzschild case, the observer cannot be positioned in an asymptotic region approximated by the Minkowski metric. To overcome this problem, we assume that the deflection angle is described by the formula \cite{usPRD}
\begin{equation}\label{formula}
\Delta\varphi(\rho_0)=\kappa_1\mathfrak{I}(\rho_0)+\kappa_2,\
\mathfrak{I}(\rho_0)=\frac{1}{\sqrt{g(\vartheta_c)}\sin{\vartheta_c}}\int_1^{\widehat{\rho}_b}\frac{d\widehat{\rho}}{\widehat{\rho}\sqrt{\widehat{\rho}^2 f_\Lambda(\rho_0)-f_\Lambda(\rho_0\widehat{\rho})}},\quad\widehat{\rho}=\frac{\rho}{\rho_0},
\end{equation}
with unknown constants $\kappa_1$ and $\kappa_2$ to be fixed so that the weak field approximation of (\ref{formula}) coincides with the weak field approximation for the Schwarzschild case in the limit of vanishing cosmological constant and acceleration parameter. The integral in (\ref{formula}) can be rewritten as 
\begin{eqnarray}
\mathfrak{I}(\rho_0)&=&\frac{1}{\sqrt{g(\vartheta_c)}\sin{\vartheta_c}}\int_1^{\widehat{\rho}_b}d\widehat{\rho}~F(\widehat{\rho};\epsilon,\mu),\quad\epsilon=\frac{1}{\rho},\quad\mu=\frac{\widehat{x}^2}{\epsilon},\\
F(\widehat{\rho};\epsilon,\mu)&=&\frac{1}{\widehat{\rho}\sqrt{\widehat{\rho}^2-1+\epsilon\left(\frac{1}{\widehat{\rho}}-\widehat{\rho}^2\right)+\mu\left(\widehat{\rho}^2-\widehat{\rho}\right)}},\quad\mu=\frac{\widehat{x}^2}{\epsilon}.
\end{eqnarray}
For the discussion on why it is possible to apply a perturbative expansion in the small parameters $\epsilon$ and $\mu$ we refer to \cite{Maha}. Therefore, let us expand $F$ as follows
\begin{equation}
F(\widehat{\rho};\epsilon,\mu)=f_0(\widehat{\rho})+f_1(\widehat{\rho})\epsilon+f_2(\widehat{\rho})\epsilon^2+f_3(\widehat{\rho})\epsilon^3+g_1(\widehat{\rho})\mu+f_4(\widehat{\rho})\epsilon^4+\mathcal{O}(\epsilon\mu)
\end{equation}
with
\begin{eqnarray}
f_0(\widehat{\rho})&=&\frac{1}{\widehat{\rho}\sqrt{\widehat{\rho}^2-1}},\quad
f_1(\widehat{\rho})=\frac{\widehat{\rho}^2+\widehat{\rho}+1}{2\widehat{\rho}^2(\widehat{\rho}+1)\sqrt{\widehat{\rho}^2-1}},\quad
f_2(\widehat{\rho})=\frac{3(\widehat{\rho}^2+\widehat{\rho}+1)^2}{8\widehat{\rho}^3(\widehat{\rho}+1)^2\sqrt{\widehat{\rho}^2-1}},\\
f_3(\widehat{\rho})&=&\frac{15(\widehat{\rho}^2+\widehat{\rho}+1)^3}{48\widehat{\rho}^4(\widehat{\rho}+1)^3\sqrt{\widehat{\rho}^2-1}},\quad
g_1(\widehat{\rho})=-\frac{1}{2(\widehat{\rho}+1)\sqrt{\widehat{\rho}^2-1}},\quad
f_4(\widehat{\rho})=\frac{105(\widehat{\rho}^2+\widehat{\rho}+1)^4}{384\widehat{\rho}^5(\widehat{\rho}+1)^4\sqrt{\widehat{\rho}^2-1}}.
\end{eqnarray}
If we take into account that 
\begin{equation}
\frac{1}{\sqrt{g(\vartheta_c)}\sin{\vartheta_c}}=1+\mathcal{O}(\epsilon\mu),
\end{equation}
and we let the integration over the functions $f_0,\cdots,f_4$ and $g_1$ to be followed by an asymptotic expansion in powers of $1/\widehat{\rho}_b$, the deflection angle in  (\ref{formula}) becomes
\begin{equation}\label{firuli}
\Delta\varphi(\rho_0)=\kappa_1\left[\mathfrak{F}_0+\mathfrak{F}_1\epsilon+\mathfrak{F}_2\epsilon^2+\mathfrak{F}_3\epsilon^3+\mathfrak{G}_1\mu+\mathfrak{F}_4\epsilon^4+\mathcal{O}(\epsilon\mu)\right]+\kappa_2
\end{equation}
where
\begin{eqnarray}
\mathfrak{F}_0&=&\frac{\pi}{2}-\frac{1}{\widehat{\rho}_b}+\mathcal{O}\left(\frac{1}{\widehat{\rho}^3_b}\right),\quad
\mathfrak{F}_1=1-\frac{1}{2\widehat{\rho}_b}+\mathcal{O}\left(\frac{1}{\widehat{\rho}^3_b}\right),\quad
\mathfrak{F}_2=\frac{15}{32}\pi-\frac{1}{2}-\frac{3}{8\widehat{\rho}_b}+\mathcal{O}\left(\frac{1}{\widehat{\rho}^3_b}\right),\\
\mathfrak{F}_3&=&\frac{61}{24}-\frac{15}{32}\pi-\frac{5}{16\widehat{\rho}_b}+\mathcal{O}\left(\frac{1}{\widehat{\rho}^3_b}\right),\quad
\mathfrak{G}_1=-\frac{1}{2}+\frac{1}{2\widehat{\rho}_b}-\frac{1}{4\widehat{\rho}_b^2}+\mathcal{O}\left(\frac{1}{\widehat{\rho}^3_b}\right),\\
\mathfrak{F}_4&=&\frac{3465}{2048}\pi-\frac{65}{16}-\frac{35}{128\widehat{\rho}_b}+\mathcal{O}\left(\frac{1}{\widehat{\rho}^3_b}\right).
\end{eqnarray}
In order to fix the unknown constants $\kappa_1$ and $\kappa_2$ in (\ref{firuli}), we observe that in the limit of $\Lambda\to 0$, the cosmological horizon $\widehat{\rho}_h\to\infty$ and therefore, we can let $\widehat{\rho}_b\to\infty$ in the above expressions. If in addition $\alpha\to 0$, equation (\ref{firuli}) must reproduce the weak deflection angle for a light ray in the Schwarzschild metric. This is the case if $\kappa_1=2$ and $\kappa_2=-\pi$. Hence, at the first order in $1/\widehat{\rho}_b$, we find that the weak deflection angle can be written as
\[
\Delta\varphi(\rho_0)=-\frac{2}{\widehat{\rho}_b}+\left(2-\frac{1}{\widehat{\rho}_b}\right)\frac{1}{\rho_0}+\left(\frac{15}{16}\pi-1-\frac{3}{4\widehat{\rho}_b}\right)\frac{1}{\rho_0^2}+\left(\frac{61}{12}-\frac{15}{16}\pi-\frac{5}{8\widehat{\rho}_b}\right)\frac{1}{\rho_0^3}+
\]
\begin{equation}\label{ecco}
4\left(-1+\frac{1}{\widehat{\rho}_b}\right)\alpha^2 M^2\rho_0+\left(\frac{3465}{1024}\pi-\frac{65}{8}-\frac{35}{64\widehat{\rho}_b}\right)\frac{1}{\rho_0^4}+\cdots.
\end{equation}
Taking into account that for a vanishing cosmological constant, we can let $\widehat{\rho}_b\to\infty$, it is straightforward to check that (\ref{ecco}) correctly reproduces the weak deflection angle formula (115) for the $C$-metric obtained in \cite{Maha}. 

Regarding the strong gravitational lensing for the metric under consideration, we first observe that it can be analyzed by the same procedure adopted by \cite{Maha} because in both cases the metrics involved admits the same family of null circular orbits. For this reason, we will not dive into the details of the derivation and we will limit us to remind the reader that one first solves the integral (\ref{formula}) in terms of an incomplete elliptic function of the first kind followed by an application of an asymptotic formula for the aforementioned elliptic function obtained by \cite{KNC} when the sine of the modular angle and the elliptic modulus both approach one. The same strategy has been already successfully used in \cite{usPRD} to derive the Schwarzschild deflection angle in the strong regime with a higher degree of precision than the corresponding formulae in \cite{Darwin,Bozza}. Without further delay, let us recall that it was found in \cite{Maha} that the deflection angle in the strong gravitational lensing regime is given by
\begin{equation}\label{SGL}
\Delta\varphi(\rho_0)=-\pi+\mathfrak{h}_1(\rho_b,\rho_c)-\mathfrak{h}_2(\rho_c)\ln{\left(\frac{\rho_0}{\rho_c}-1\right)}-\mathfrak{h}_3(\rho_c)(\rho_0-\rho_c)+\mathcal{O}(\rho_0-\rho_c)^2
\end{equation}
with
\begin{eqnarray}
\mathfrak{h}_1(\rho_b,\rho_c)&=&\mathfrak{h}_2(\rho_c)\ln{\frac{8(3-\rho_c)(\rho_b-\rho_c)}{\left[\sqrt{\rho_b(3-\rho_c)}+\sqrt{\rho_c+2\rho_b-\rho_b\rho_c}\right]^2}},\\
\mathfrak{h}_2(\rho_c)&=&\frac{3\sqrt{6}\rho_c}{(3+\rho_c)\sqrt{3-\rho_c}},\quad
\mathfrak{h}_3(\rho_c)=\frac{3\sqrt{6}(2\rho_c-7)}{2(\rho_c+3)(3-\rho_c)^{3/2}}.
\end{eqnarray}
A validity check of (\ref{SGL}) was already run in \cite{Maha} where it has been verified that (\ref{SGL}) correctly reproduces the corresponding strong lensing formula in the Schwarzschild case as given in \cite{usPRD,Bozza} when $\alpha\to 0$. The main difference between (143) in \cite{Maha} and (\ref{SGL}) revolves around $\rho_b$, i.e. the position of the observer, which in the present case is bounded from above by the cosmological horizon. It is interesting to observe that, in general, the above formula does not depend on $\Lambda$. However, a dependence on the cosmological constant emerges only in the case the observer is placed very close to the cosmological horizon as it was already pointed out by \cite{usPRD} in the context of the Schwarzschild-de Sitter metric.

\section{Conclusions}

In this paper, we have focussed on the classical connection between light and gravity, more precisely, the bending of light in a gravitational field and its lensing. For the $C$-metric with positive cosmological constant, we showed that the effective potential for a massless particle exhibits a saddle point. If on one hand a local maximum in the effective potential corresponds to an unstable null circular orbit, on the other hand, the presence of a saddle point leads to a more challenging classification problem which needs a careful scrutiny. By means of a Jacobi analysis we showed that the light-like circular geodesics associated to the aforementioned saddle point are unstable. Furthermore, we constructed the impact parameter for the light scattering in the dS $C$-metric and showed that the Sachs scalars do not depend on the cosmological constant, hence they cannot be used to optically discriminate among $C$- and $C$- black holes with $\Lambda$. This obliged us to probe into the weak and strong gravitational lensing for which we computed the corresponding deflection angle in terms of the distance of closest approach and the position of the observer. Our results reveal that corrections of the cosmological constant appear only in the case the observer is located close to the cosmological horizon.

\end{document}